\documentclass[prb,aps,showpacs,twocolumn,floatfix,citeautoscript,superscriptaddress]{revtex4-1}
\usepackage{amsmath,amssymb,physics,color,array}
\usepackage{graphicx}
\usepackage{epstopdf}
\usepackage{color}
\begin{document}

\title{Nonadiabatic Josephson current pumping by microwave irradiation}

\author{B. Venitucci}
\affiliation{Centre National de la Recherche Scientifique, Institut NEEL, F-38042 Grenoble Cedex 9, France}
\affiliation{Universit\'e Grenoble-Alpes, Institut NEEL, F-38042 Grenoble Cedex 9, France}
\email[Present address: ]{Universit\'e Grenoble-Alpes and Commissariat à l'Energie Atomique, INAC-MEM,  F-38000 Grenoble, France}
\author{D. Feinberg}
\affiliation{Centre National de la Recherche Scientifique, Institut NEEL, F-38042 Grenoble Cedex 9, France}
\affiliation{Universit\'e Grenoble-Alpes, Institut NEEL, F-38042 Grenoble Cedex 9, France}
\author{R. M\'elin}
\affiliation{Centre National de la Recherche Scientifique, Institut NEEL, F-38042 Grenoble Cedex 9, France}
\affiliation{Universit\'e Grenoble-Alpes, Institut NEEL, F-38042 Grenoble Cedex 9, France}
\author{B. Dou\c{c}ot}
\affiliation{Laboratoire de Physique Th\'eorique et des Hautes Energies,
  CNRS UMR 7589, Universit\'es Paris 6 et 7, 4 Place Jussieu, 75252 Paris
  Cedex 05}
  
\begin{abstract}
Irradiating a Josephson junction with microwaves can operate not only
on the amplitude but also on the phase of the Josephson current. This
requires breaking time inversion symmetry,
which is achieved by introducing a
  phase lapse between the microwave components acting on the two†
sides of the junction. General symmetry arguments
and the solution of a specific single level quantum
dot model show that this induces chirality in the
Cooper pair dynamics, due to the topology of the Andreev bound state wavefunction. Another essential condition is to
break electron-hole symmetry within the junction. A shift of the
current-phase relation is obtained, which is controllable in sign and amplitude with the
microwave phase and an electrostatic gate, thus
  producing a ``chiral'' Josephson transistor. The dot model is solved in
the infinite gap limit by Floquet theory and in the general case
with Keldysh nonequilibrium Green's functions. The
chiral current is nonadiabatic: it is
extremal and changes sign close to resonant chiral
transitions between the Andreev bound states.
\end{abstract}


\maketitle

\section{Introduction}
Microwave irradiation has always been a privileged tool to analyze Josephson
junctions\cite{Barone}. Photon-assisted Cooper pair transport has been
observed in biased junctions
\cite{Dayem,Tien-Gordon}. Shapiro steps reveal
synchronization of the Josephson ac oscillations to the
microwave excitation\cite{Shapiro}. In transparent junctions such as quantum
point contacts and in junctions made of a quantum dot with a few
levels, the Josephson properties are governed by a discrete set of
Andreev bound states (ABSs)\cite{ABS}. In absence
of constant bias, nonadiabatic behavior is present
even at low irradiation when the microwave frequency (or its
harmonics) matches the ABS spacing\cite{Bergeret1, Bergeret2}, causing
a sharp decrease of the current amplitude. Resonant
microwave can be used as a spectroscopic probe of the
ABS dispersion with the phase difference applied on the
junction\cite{Bretheau}.

Josephson currents can be induced either by a magnetic flux of by driving a DC current through a junction. Here we propose a third way 
of inducing a Josephson current. The main result of this work is that microwave
radiation can pump a Josephson current in absence
of applied superconducting phase difference and
at zero average bias voltage. The
required breaking of time inversion symmetry
originates from the phase of the microwave
radiation, applied as an oscillating voltage
$v_j(t)$ on each side $j=1,\,2$ of the junction. A
nontrivial phase difference $\chi=\chi_1-\chi_2$ is
  included in microwave voltage amplitudes $v_j(t)=-v_j\sin(\Omega
t+\chi_j)$, which introduces chirality in the system. This can be achieved for instance by using a common
microwave line bifurcating into two branches, one containing a delay
line. A nonzero current appears for
$\chi\neq0,\,\pi$, made of Cooper
pairs pumped through the junction. Another and less obvious condition is to break
charge conjugation symmetry, e.g. electron-hole
symmetry in the junction. As shown below, this can be related to general symmetry considerations. An especially interesting situation 
is met if the junction is made with a gated quantum dot. The resulting description, 
adopted in this paper, considers a single noninteracting level, leading to a pair of Andreev bound states (ABS). A simplified ``toy-model'' description 
is obtained in the infinite-gap model (IGM), yielding a periodically driven two-level system that can be solved with Floquet formalism. This model captures the essential physics. 
The general case, involving coupling to the quasiparticle continuum states, is analyzed using nonequilibrium Keldysh Green's functions. 

This effect pertains to the wide class of quantum pumping phenomena\cite{Thouless,Niu}. Yet, it is remarkable that 
it disappears in the adiabatic limit, e.g. when the quantum state of the junction is adiabatically modulated by the
microwave radiation. Comparing the microwave frequency to the energy splitting between the phase-dependent ABS, three regimes are obtained : i) a low-frequency regime,
 which can be described within the IGM by Thouless' argument\cite{Thouless}, to lowest order in nonadiabaticity; ii) a high-frequency regime, easily solvable analytically within the IGM; iii) a resonant regime,
 which generalizes the current anomalies found by Bergeret et al.\cite{Bergeret1, Bergeret2}, and amenable to a rotating-wave-approximation (RWA) solution in the IGM. 
 Special attention is paid to the cases of superconducting phase differences $\varphi=0,\pi$ where the Josephson current is exclusively due to quantum pumping. 
 For simplicity, this current will be hereafter called ``chiral current'', not to be confused with the oriented Josephson current created by a magnetic flux piercing a ring. 

One must emphasize that the present pumping mechanism differs from other ones studied in Josephson junctions, such as involving Coulomb blockade\cite{Pothier},
 biased junctions\cite{Kopnin}, or a biased SQUID pumping a normal
current\cite{Giazotto,Russo}. Our results also offer a new
way to create a highly tunable
  $\varphi_0$-junction with shifted current-phase
relation (CPR)\cite{Bulaevskii,Krive,Egger,Reynoso,Buzdin,Liu,Sickinger,Yokohama,Szombati}. 

The plan of this work is as follows. After the Introduction (Section 1), Section 2 presents the model and a general analysis of the underlying symmetries. Section 3 contains the Floquet solution of the 
infinite-gap model : low and high frequency, and close to a resonance (RWA). Section 4 provides the Keldysh solution of the full model and compares it to that of the IGM. Section 5 demonstrates a mapping of the IGM onto a 
tight-binding lattice chain model, emphasizing the relation between the problem considered in this work and some driven lattice models. Section 6 concludes the paper. 

\section{The model and its symmetries}
The
time-dependent phases deriving from the applied microwave voltages $v_1(t)=-v_1\sin(\Omega t+\frac{\chi}{2})$, $v_1(t)=-v_2\sin(\Omega t-\frac{\chi}{2})$ are defined as:
\begin{equation}
  \label{eq:1}
 \varphi_1=\frac{\varphi}{2}+b_1\cos(\Omega
 t+\frac{\chi}{2}),\;\varphi_2=-\frac{\varphi}{2}+b_2\cos(\Omega
 t-\frac{\chi}{2})
 \end{equation}
 with $b_j=\frac{2ev_j}{\hbar \Omega}>0$. To take a simple example, let us first consider a tunnel junction and 
 work in the adiabatic approximation, e.g. plugging the phase dependences into
 the equilibrium current-phase relation $I(\varphi_1,\varphi_2)=I_c\sin(\varphi_1-\varphi_2)$. Using
 Bessel function expansion, one finds that the microwave radiation only modifies
the amplitude of the critical current according to :
\begin{equation}
\langle
I \rangle_{dc}=I_0[ J_0(b_1) J_0(b_2) + 2\sum_{n>0} J_n(b_1) J_n(b_2)
  \cos(n\chi)]\,\sin\varphi.
\end{equation}
We show below that going beyond the adiabatic regime in a quantum dot junction 
makes the microwave affect not only the amplitude but also the phase of the Josephson current.

A specific model of gate-tunable quantum dot
Josephson junction is considered now, which bears a
single relevant level in front of the energy gap of the
superconductors $j=1,2$. Neglecting Coulomb interactions,
the Hamiltonian is the following:
\begin{eqnarray}
\label{Hamilt}
H&=&\sum_{kj\sigma} \xi_{kj} c^{\dagger}_{kj\sigma} c_{kj\sigma}\,+\,\Delta \sum_{kj} (c^{\dagger}_{kj\uparrow} c^{\dagger}_{-kj\downarrow}+H.c.)\\
\nonumber
&+&\varepsilon_0 \sum_{\sigma}d^{\dagger}_{\sigma} d_{\sigma}\,+\,\sum_{kj,\sigma} t_j \big[c^{\dagger}_{kj\sigma} d_{\sigma}\,e^{i\varphi_j(t)/2}+H. c. \big]\;,
\end{eqnarray}
with $\xi_{k\sigma}=\varepsilon_{k\sigma}-\mu$. The potentials
$ev_j(t)$ and the phases in the pairing terms have been gauged
away to appear in the tunneling terms towards or
from the dot. 

Unlike the case $v_j=0$, the physical
properties do not depend in general solely on the
phase difference
$\varphi(t)=\varphi_1(t)-\varphi_2(t)$, as seen by
performing the gauge transformation
$U=\exp\big(-i\nu\varphi_2(t)\big)$ (defining $\nu=\frac{1}{2}\sum_{\sigma}
d^{\dagger}_{\sigma} d_{\sigma}$). In the presence of
  time-dependent phases, this transformation correctly eliminates the
phase $\varphi_2(t)$ but it also yields a
time-dependent gate voltage on the dot. Actually, the
Hamiltonian~(\ref{Hamilt}) does depend on two independent
time-dependent fields, e.g. it can lead to quantum pumping if the
phase lapse $\chi$ is different from $0$ or $\pi$.

We now investigate the symmetries of the full
Hamiltonian~(\ref{Hamilt}), first in absence of microwave
  radiation. Eq.~(\ref{Hamilt}) is then parameterized by the phase
  $\varphi$ and the dot energy $\varepsilon_0$. Time inversion
  $\mathcal{T}$ and charge conjugation $\mathcal{C}$
  act on the fermion operators as
  $\mathcal{T}c_{k\sigma}\mathcal{T}^{-1}=-\sigma c_{-k,-\sigma}$
  ($\sigma=\pm$) and
  $\mathcal{C}c_{k\sigma}\mathcal{C}^{-1}=c^{\dagger}_{k\sigma}$. Using
  antilinearity of $\mathcal{T}$ leads to $\mathcal{T} H(\varphi ,
  \varepsilon_0) \mathcal{T}^{-1} = H(-\varphi , \varepsilon_0)$. The
  current operator $\hat{J}(\varphi)=\frac{2e}{\hbar}\frac{\partial
    H}{\partial \varphi}$ transforms according to
  $-\hat{J}(-\varphi)$, yielding the usual symmetry:
  \begin{equation}
  \label{Josephsonsymmetry}
  \langle
  \hat{J}(-\varphi,\varepsilon_0)\rangle=-\langle
  \hat{J}(\varphi,\varepsilon_0)\rangle.
  \end{equation}
  On the other hand, charge
  conjugation applied to Eq.~(\ref{Hamilt}) turns $\xi_{ki\sigma},\,
  t_i,\,\varepsilon_0$ and $\varphi_i$ into $-\xi_{ki\sigma},\,
  -t_i,\,-\varepsilon_0$ and $-\varphi_i$. Owing to symmetries of the
  current with $\xi$ and $t_i$, applying $\mathcal{CT}$ leads to:
  \begin{equation}
  \label{Josephsontransistor}
  \langle \hat{J}(\varphi,-\varepsilon_0)\rangle=\langle
  \hat{J}(\varphi,\varepsilon_0)\rangle, 
  \end{equation}
  a well-known symmetry of the so-called Josephson transistor\cite{JTransistor}. 
  
  Let us now show that this last symmetry is
  broken by the chiral phase $\chi$. The symmetry operators $\mathcal{T}$ and
$\mathcal{C}$ are applied to the time-dependent
Hamiltonian $H(\varphi_j(t))$. Then
$\mathcal{T}H(\varphi_j(t),\varepsilon_0)\mathcal{T}^{-1}=H(-\varphi_j(-t),\varepsilon_0)$
and $-\varphi_j(-t)=-\varphi_j+b_j\cos(\Omega t'-\chi_j)$ with
$\varphi_j=\pm\frac{\varphi}{2},\chi_j=\pm\frac{\chi}{2}$,
$t'=t+\frac{\pi}{\Omega}$. The latter time translation leaves the time-averaged quantities unchanged, therefore, 
following the same reasoning as above, one obtains for the DC component of the current:
 \begin{equation}
 \label{sym1}
  \langle\hat{J}(-\varphi,-\chi,\varepsilon_0)\rangle_{dc}=-\langle \hat{J}(\varphi,\chi,\varepsilon_0)\rangle_{dc}
  .
 \end{equation}
  Similarly, applying $\mathcal{C}$ leaves $\chi$
  unchanged, leading to
 \begin{equation}
  \label{sym2}
 \langle \hat{J}(-\varphi,\chi,-\varepsilon_0)\rangle_{dc}=-\langle \hat{J}(\varphi,\chi,\varepsilon_0)\rangle_{dc}
 .
 \end{equation}
  Eq. (\ref{sym1}) shows that time inversion operates both on $\chi$ and $\varphi$, allowing in principle to
  generate a nonzero ``chiral'' current with $\varphi=0,\,\pi$ but $\chi\neq0,\,\pi$. Let us define these currents by $J_{chir,0/\pi}(\chi,\varepsilon_0)=\langle \hat{J}(0/\pi,\chi,\varepsilon_0)\rangle_{dc}$. 
  Equations (\ref{sym1},\ref{sym2}) lead to:
  \begin{eqnarray}
  \label{Jsymchiral}
  J_{chir,0/\pi}(-\chi,\varepsilon_0)&=&-J_{chir,0/\pi}(\chi,\varepsilon_0)\\ 
  \label{Jtranschiral}
  J_{chir,0/\pi}(\chi,-\varepsilon_0)&=&-J_{chir,0/\pi}(\chi,\varepsilon_0)
  \end{eqnarray}
  Therefore the chiral current not only changes sign with $\chi$ - Equation (\ref{Jsymchiral}) is similar to Equation (\ref{Josephsonsymmetry}) - but Equation (\ref{Jtranschiral}) show that it also changes sign with $\varepsilon_0$, contrarily to the current
  generated by $\varphi$ only (Equation (\ref{Josephsontransistor})). This nontrivial connection between time
  inversion and electron-hole symmetries is a fingerprint of this
  chiral Josephson current. Notice that a normal
  current pumped through a gated quantum dot is also found to change
  sign with the gate voltage\cite{Kouwenhoven}.

\section{The infinite-gap limit: Floquet analysis}
\subsection{The Hamiltonian}
Let us first consider the infinite-gap model (IGM) limit
  \cite{Jonckheere}. The subspaces of even and odd number states on the dot become decoupled and 
  only the even states may mediate a Josephson coupling. Within the 
  even number space, a pseudospin mapping of the empty and doubly
occupied states is defined as:
$\hat{\tau}_{+}=d^{\dagger}_{\uparrow}d^{\dagger}_{\downarrow}$ and
$\hat{\tau}_{z}=2d^{\dagger}_{\uparrow}d^{\dagger}_{\downarrow}d_{\downarrow}d_{\uparrow}-1$,
yielding a driven two-level system:
\begin{equation}
\label{twolevel}
H_{\infty}(t)=\sum_{j=1,2} \gamma_j\Big[e^{-i\varphi_j(t)}\hat{\tau}_{+}+e^{i\varphi_j(t)}\hat{\tau}_{-}\Big]+\varepsilon_0 \hat{\tau}_z\;,
\end{equation}
where $\gamma_j=\pi \nu(0)t_j^2$ is the pair hopping
amplitude [$\nu(0)$ is the metallic density of
  states]. 
  
This model can be solved with Floquet theory\cite{Grifoni}. The wavefunction $\Psi(t)$ evolves according to
$i\dot{\Psi}(t)=H_\infty(t)\Psi(t)$ (taking $\hbar=1$). For the
$T$-periodic Hamiltonian $H_\infty(t)$ ($T=2\pi/\Omega$), there exists a set
of Floquet pseudo-energies ($\epsilon_\alpha$) and a periodic basis of
wave-functions $(\phi_\alpha)$ with period $T$. A basis solution $\Psi_\alpha(t)$ can be written as $\Psi_\alpha(t)=e^{-i\epsilon_\alpha t
}\phi_\alpha(t)$.

Fourier expansion of $H_\infty(t)$ and $\phi_\alpha(t)$ gives:
\begin{equation}
H_\infty(t) = \sum_{n=-\infty}^{+\infty} \tilde{H}_{\infty n} e^{in\Omega t} \quad\textrm{,}\quad 
\phi_\alpha(t) = \sum_{n=-\infty}^{+\infty} \tilde{\phi}_{\alpha  n} e^{in\Omega t},
\label{decomp_serie_fourier}
\end{equation}
where, defining $\gamma_{i,n}=\gamma_iJ_n(b_i)(i=1,2)$
\begin{eqnarray}
\nonumber
\tilde{H}_{\infty,0} &=&
\big(\gamma_{1,0}e^{-i\frac{\varphi}{2}}+\gamma_{2,0}e^{i\frac{\varphi}{2}}\big)\,\tau_{+}\\
&+&\big(\gamma_{1,0}e^{i\frac{\varphi}{2}}+\gamma_{2,0}e^{-i\frac{\varphi}{2}}\big)\,\tau_{-}+\varepsilon_0\tau_z,
\end{eqnarray} 
and for all integers $n\neq0$:
\begin{eqnarray}
\nonumber
\tilde{H}_{\infty,n} &=&
[(- i)^n\big(\gamma_{1,n}{e}^{- {i}\frac{\varphi}{2}} {e}^{ {i}n\frac{\chi}{2}}+\gamma_{2,n}  {e}^{ {i}\frac{\varphi}{2}}  {e}^{- {i}n\frac{\chi}{2}}]\,\tau_{+} \\
 &+&[(i^n\big(\gamma_{1,n}{e}^{i\frac{\varphi}{2}} {e}^{ {i}n\frac{\chi}{2}}+\gamma_{2,n}  {e}^{ -{i}\frac{\varphi}{2}}  {e}^{- {i}n\frac{\chi}{2}}]\,\tau_{-}.
\end{eqnarray}  

The Fourier series defined in
Eq. \eqref{decomp_serie_fourier} leads to
\begin{equation}
  \sum_{m=-\infty}^\infty\tilde{H}^{FL}_{\infty nm}\phi_{\alpha m}=\epsilon_\alpha\phi_{\alpha n}
  ,
\label{shro_harm}
\end{equation}
where 
\begin{equation}
\tilde{H}^{FL}_{\infty nm}= \tilde{H}_{\infty,
  n-m}+n\Omega\delta_{nm}.
 \end{equation}
Noting $\tilde{\phi}_\alpha=
\begin{pmatrix}
\hdots ,
\tilde{\phi}_{\alpha -1},
\tilde{\phi}_{\alpha 0},
\tilde{\phi}_{\alpha 1},
\hdots 
\end{pmatrix}
,
$
Eq. \eqref{shro_harm} can be cast in matrix form:
\begin{equation}
\tilde{H}^{FL}_\infty\tilde{\phi}_\alpha = \epsilon_\alpha\tilde{\phi}_\alpha.
\label{eigenvalue_floquet}
\end{equation}

The DC-current $I_\alpha$ associated to state
$\ket{\Psi_\alpha(t)}$ is
defined as $I_\alpha=2e\overline{\bra{\Psi_\alpha(t)}\frac{\partial
    H_\infty(t)}{\partial \varphi}\ket{\Psi_\alpha(t)}}$ where
$\overline{f(t)}=\frac{1}{T}\int_0^Tf(t) {dt}$. Eq.~\eqref{decomp_serie_fourier} leads to
\begin{equation}
\overline{\bra{\Psi_\alpha(t)}\frac{\partial H_\infty(t)}{\partial \varphi}\ket{\Psi_\alpha(t)}} =\bra{\tilde{\phi}_\alpha}\frac{\partial \tilde{H}^{FL}_\infty}{\partial \varphi}\ket{\tilde{\phi}_\alpha}
.
\end{equation}
According to Hellman-Feynman theorem,
$\bra{\tilde{\phi}_\alpha}\frac{\partial \tilde{H}^{FL}}{\partial
  \varphi}\ket{\tilde{\phi}_\alpha} = \frac{\partial
  \epsilon_\alpha}{\partial \varphi}$. Therefore, the DC-current
$I_\alpha$ in  Floquet eigenstate becomes
\begin{equation}
I_\alpha(\varphi,\chi,\epsilon_0) =2e\frac{\partial \epsilon_\alpha}{\partial \varphi}(\varphi,\chi,\epsilon_0)
\label{current_floquet}
,
\end{equation}
generalizing the Josephson equation to a Floquet state.
The DC-current $I_\alpha$ can be
calculated from Eq. \eqref{current_floquet} once the pseudo-energy
$\epsilon_\alpha$ are known from Eq. \eqref{eigenvalue_floquet}.

\subsection{Low-frequency limit}
In this section, the microwave frequency $\Omega$ is supposed to be
small compared to the other relevant energies. The quasiadiabatic
approximation can then be used.

Let us call $(\ket{\Psi_+(t)},\ket{\Psi_-(t)})$, the instantaneous
basis of the system, such that
$H_\infty(t)\ket{\Psi_\pm(t)}=E_\pm(t)\ket{\Psi_\pm(t)}$, where
\begin{equation}
E_\pm(t)=\pm\sqrt{\epsilon_0^2+\gamma_1^2+\gamma_2^2+2\gamma_1\gamma_2\cos(\varphi_1(t)-\varphi_2(t))}.
\end{equation}
with $\varphi_{1,2}(t)$ given by Equation (\ref{eq:1}). 
Each wavefunction
$\ket{\Psi(t)}$ can be decomposed on this basis : $\ket{\Psi(t)}
=c_+(t)\ket{\Psi_+(t)}+c_-(t)\ket{\Psi_-(t)}$. Following Thouless
\cite{Thouless} (see also Xiao \cite{Xiao})and choosing
$\ket{\Psi(0)}=\ket{\Psi_+(0)}$, we have
\begin{equation}
\ket{\Psi(t)}=\ket{\Psi_+(t)}
-i\ket{\Psi_-(t)}\frac{\langle \Psi_-(t)|\frac{\partial}{\partial t}\Psi_+(t)\rangle}{E_+(t)-E_{-}(t)}
.
\label{state_quasi_adiabatic}
\end{equation}

The DC-current of the state $\ket{\Psi(t)}$ is
$I=2e\overline{\bra{\Psi(t)}\frac{\partial H(t)}{\partial
    \varphi}\ket{\Psi(t)}}$ and one finds
\begin{equation}
  I=2e( \overline{\frac{\partial E_+}{\partial \varphi}}-\overline{C^+_{\varphi,t}}),
  \label{eq:XX}
\end{equation}
where 
\begin{equation}
C^+_{\varphi,t} = i\left(\langle{\frac{ \partial \Psi_+}{
    \partial\varphi}|\frac{\partial\Psi_+}{\partial t}}\rangle-\langle{\frac{\partial\Psi_+}{
    \partial t}|\frac{\partial\Psi_+}{\partial\varphi}}\rangle\right)
 \end{equation}
    is the Berry
curvature in the $(\Omega t,\varphi)$ variables.  The
first term in Eq.~(\ref{eq:XX}) is the adiabatic contribution and the
second one is the first nonadiabatic correction.  It is straightforward to check that if
$\varphi=0,\pi$ the adiabatic current  $2e\frac{\partial
  E_+}{\partial
  \varphi}=\frac{-\gamma_1\gamma_2\sin(\varphi_1(t)-\varphi_2(t))}{|E_{-}(t)|}$ has a zero time average, 
  whatever the relative phase $\chi$ of the microwave fields. Actually, the adiabatic current is a function of 
  $\varphi(t)=\varphi_1(t)-\varphi_2(t)=\varphi+b_1\cos(\Omega t+\frac{\chi}{2})-b_2\cos(\Omega t-\frac{\chi}{2})$, 
  which verifies $\varphi(t+\frac{T}{2})=-\varphi(t)$ and thus has a zero integral on the interval $[0,T]$\cite{footnote}. 
 Therefore the chiral current at $\varphi=0,\,\pi$ is
intrinsically nonadiabatic, and
\begin{equation}
I_{chir,0/\pi}= -2e\,\overline{C^+_{\varphi,t}}\lvert_{\varphi=0/\pi}
\label{current_berry_curvature}
.
\end{equation}

This justifies the word ``chiral'' used throughout the paper : chirality is related to the properties of the ABS wavefunctions. 
The Berry curvature is expressed by rewriting the Hamiltonian as
$H_\infty(t)= \vec{h}(t).\vec{\sigma}$ with:
\begin{eqnarray}
\label{effective_field}
\left\{
\begin{array}{ll}
\vec{h}(t)= \Big(
\frac{\Gamma(t)+\Gamma^*(t)}{2} ,
i\frac{\Gamma(t)-\Gamma^*(t)}{2} ,
\epsilon_0\Big)^t\\
\Gamma(t)=\sum_{j=1,2} \gamma_j e^{-i\varphi_j(t)}\\
\vec{\sigma} =(\sigma_x,\sigma_y,\sigma_z)
\end{array}
,
\right.
\end{eqnarray}
 where $|\vec{h}(t)|=|E_\pm(t)|$. Expressing
each of the three terms separately leads to
\begin{equation}
\frac{1}{|\vec{h}|^3}\vec{h}.(\partial_\varphi \vec{h}\times\partial_t \vec{h})=-2C^+_{\varphi,t}
\label{h_berry}
.
\end{equation}
Eqs. \eqref{current_berry_curvature} and \eqref{h_berry} then yield\cite{HasanKane}:
\begin{equation}
\label{nonadiabatic_current}
I_{\varphi=0}=\frac{e\Omega}{h}\frac{1}{|\vec{h}|^3}\int_0^{2\pi} \vec{h}.(\partial_\varphi\vec{h}\times\partial_{(\Omega t)}\vec{h}) {d(\Omega t)}
\end{equation}

Interestingly, this results in:
\begin{equation}
I_{\varphi=0}=-\frac{e\Omega}{2\pi} \frac{\partial}{\partial \chi}\left(\int_0^{2\pi} \frac{\epsilon_0}{|E_\pm(\Omega t)|} {d(\Omega t)}\right)
\label{calcul_adiab}
.
\end{equation}
This equation shows an intriguing formal similarity between the pumped Josephson current, as a function of the microwave phase $\chi$, and the equilibrium Josephson current, expressed as the derivative of the ABS energy with respect to the superconducting phase. Yet in Eq. (\ref{calcul_adiab}), the ABS energy is replaced by its inverse. 
 
\subsection{High-frequency limit}
Perturbation theory can be used in the case of large microwave
frequency $\Omega$ and small amplitudes $b_1$, $b_2$ of the
microwave\cite{Aoki}. The Floquet Hamiltonian is split as a diagonal and
nondiagonal part: $H_\infty^{FL} = H^{FL}_{diag}+\rho\, h$
where $ H^{FL}_{diag}$ is diagonal, and $\rho \, h$ is non
diagonal ($\rho<<1$). The resulting effective Hamiltonian takes the form
\begin{equation}
H^{FL}_{\infty,eff} =  {e}^{-i\rho S}H_\infty^{FL} {e}^{i\rho S}.
\end{equation}
The matrix $S$ is chosen according to
\begin{equation}
\bra{\alpha}S\ket{\beta}= -\frac{\bra{\alpha}h\ket{\beta}}{\epsilon_\alpha-\epsilon_\beta},
\end{equation}
where $\ket{\alpha}=\ket{\pm,p}$, $\ket{\beta}=\ket{\pm,q}$ (with $p\neq q$ integers) are eigenstates of $H^{FL}_{diag}$ with eigenvalues $\epsilon_r=\pm \sqrt{\varepsilon_0^2+\left|\tilde{z}\right|^2}+r\Omega$ and $\tilde{z}=\gamma_1 e^{-i\varphi/2}+\gamma_2 e^{i\varphi/2}$ ($r=p,q$).

Expanding the effective Hamiltonian in powers of $b$ leads to
$H^{FL}_{\infty,eff} \simeq
H^{FL}_{\infty,diag}+\frac{\rho^2}{2}[h,S]$. The effective Hamiltonian becomes\cite{Aoki} :
\begin{equation}
H^{FL}_{\infty,eff} = H^{FL}_{\infty,diag}+\frac{1}{\Omega}[\tilde{H}_{\infty,1},\tilde{H}_{\infty,-1}]1_{N\times N}
\label{h_fl_eff}
\end{equation}
where $\tilde{H}_{\infty, 1}$ (respectively $\tilde{H}_{\infty,-1}$)
is the $1$st harmonics (respectively the $-1$th) of the periodic
Hamiltonian $H_{\infty(t)}$, which
leads to:
\begin{equation}
\frac{1}{\Omega}[\tilde{H}_{\infty
    1},\tilde{H}_{\infty-1}]=\frac{1}{\Omega}\gamma_1\gamma_2 b_1
b_2\sin\varphi\sin\chi\,\sigma_z .
\end{equation}
The effective
Hamiltonian $H^{FL}_{\infty,eff}$ is block diagonal and only the zeroth harmonic is considered by periodicity of the energy
spectrum. Denoting $H_{\infty,eff}=
H^{FL}_{\infty,eff,0}$ leads to
\begin{equation}
H_{\infty,eff} = 
\big(\gamma_{1}e^{-i\frac{\varphi}{2}}+\gamma_{2}e^{i\frac{\varphi}{2}}\big)\,\tau_{+}+\big(\gamma_{1}e^{i\frac{\varphi}{2}}+\gamma_{2}e^{-i\frac{\varphi}{2}}\big)\,\tau_{-}+\tilde{\varepsilon}_0\tau_z,
\end{equation}
with the dot level renormalized by the microwave:
\begin{equation}
\tilde{\epsilon}_0 = \epsilon_0+\frac{1}{\Omega}\gamma_1\gamma_2 b_1  b_2\sin\varphi\sin\chi.
\end{equation}

The eigenvalues $H_{\infty,eff}$ are:
\begin{equation}
\tilde{E}_{0,\pm}=\pm\sqrt{\tilde{\epsilon}_0^2(\varphi)+\gamma_1^2+\gamma_2^2+2\gamma_1\gamma_2\cos\varphi}
\label{energie_propre_pert}
.
\end{equation}
Using the fact that $I=2e\pdv{\tilde{E}_{0,\pm}}{\varphi}$, the chiral current at $\varphi=0$ $(\zeta=1)$ or $\varphi=\pi$ $(\zeta=-1)$ is given by:
\begin{equation}
I_{chir,0/\pi} = 2e\frac{\epsilon_0\gamma_1\gamma_2 b_1 b_2\sin\chi}{\Omega\sqrt{\epsilon_0^2+(\gamma_1+\zeta\gamma_2)^2}}
\label{courant_pert}
.
\end{equation}
Eq.~(\ref{courant_pert}) provides evidence for a
Josephson current induced solely  by the
chiral phase $\chi$. This current is quadratic in the microwave amplitude and it 
changes sign with the dot level energy
$\varepsilon_0$, in agreement with the general symmetry
relations discussed above. If $\varphi=\pi$ and for a symmetric
  junction, it shows a jump at $\varepsilon_0=0$. 

\subsection{Solution close to resonance}
We use the same method as Bergeret {\it et al.} \cite{Bergeret2} to
find the analytical current close to resonance {\it e.g.}  when the
microwave frequency matches the ABS spacing. Let us consider the
infinite-gap Hamiltonian:
\begin{equation}
H_\infty(t)=
\begin{pmatrix}
\epsilon_0 & \Gamma(t)\\
\Gamma^*(t) & -\epsilon_0
\end{pmatrix}
\label{H_infty_rwa}
,
\end{equation}
with $\Gamma(t)=\sum_{j=1,2}\gamma_je^{-i\varphi_j(t)}=|\Gamma(t)|e^{i\beta(t)}$. The
instantaneous eigenvalues of Eq.~(\ref{H_infty_rwa}) take the form
$E_\pm(t)=\pm E_A(t)=\pm\sqrt{\epsilon_0^2+|\Gamma(t)|^2}$, and the
orthonormal eigenvectors are the following:
\begin{equation}
\ket{\Psi_+}=\begin{pmatrix}
\cos\frac{\theta}{2}e^{i\beta}\\
\sin\frac{\theta}{2}
\end{pmatrix}
\quad
\ket{\Psi_-}=\begin{pmatrix}
-\sin\frac{\theta}{2}e^{i\beta}e^{i\alpha}\\
\cos\frac{\theta}{2}e^{i\alpha}
\end{pmatrix}
,
\label{eigenvectors}
\end{equation}
where $\cos[\theta(t)]=\frac{\epsilon_0}{E_A(t)}$, $\sin[\theta(t)]=\frac{|\Gamma(t)|}{E_A(t)}$, $\alpha$ is an arbitrary phase and $H_\infty(t)\ket{\Psi_\pm}=\pm E_A(t)\ket{\Psi_\pm}$. 

Let us now rotate to the instantaneous basis
$\Big(\ket{\Psi_+},\ket{\Psi_-}\Big)$. In this new basis,
$\ket{\tilde{\Psi}}$ is related to $\ket{{\Psi}}$ in the old basis by
$\ket{\tilde{\Psi}}=U\ket{\Psi}$ where U is a unitary matrix, obtained from Eq. \eqref{eigenvectors}:
\begin{equation}
U^\dagger=
\begin{pmatrix}
\cos\frac{\theta}{2}e^{i\beta} & -\sin\frac{\theta}{2}e^{i\beta}e^{i\alpha}\\
\sin\frac{\theta}{2} &
\cos\frac{\theta}{2}e^{i\alpha}
\end{pmatrix}
.
\end{equation}
The gauge $\alpha(t)$ eliminates the $\hat{\tau}_x$ term in the Hamiltonian $\hat{H}_A$. It must satisfy
\begin{equation}
\dot{\beta}\sin\theta \cos\alpha+\dot{\theta}\sin\alpha=0.
\end{equation}

The new Hamiltonian associated to $\ket{\tilde{\Psi}}$ is $\hat{H}_A=UHU^\dagger+i\dv{U}{t}U^\dagger$. It is given by:
\begin{equation}
\hat{H}_A(t)=[E_A(t)+\dot{\beta}\cos^2\frac{\theta}{2}-\frac{\dot{\beta}+\dot{\alpha}}{2}]\hat{\tau}_z-\frac{\dot{\theta}}{2\cos\alpha}\hat{\tau}_y+\frac{\dot{\beta}+\dot{\alpha}}{2}
.
\end{equation}

The current operator is defined as
$I_\infty=2e\pdv{H_\infty}{\varphi}$. From now on, we choose a symmetric
junction ($\gamma_1=\gamma_2=\gamma_0$) and symmetrical microwave amplitudes ($b_1=b_2=b$). The angle $\beta$ does not depend anymore
on $\varphi$. Using Eq.~\eqref{H_infty_rwa}, the current operator
takes the form $I_\infty =
2e\,\pdv{|\Gamma(t)|}{\varphi}[\cos\beta\,\hat{\tau}_x-\sin\beta\,\hat{\tau}_y]$. In
the new basis, the current operator becomes $\hat{I}_A=UI_\infty
U^\dagger$, thus:
\begin{equation}
\hat{I}_A=2e\pdv{|\Gamma(t)|}{\varphi}[\cos\theta\,\hat{\tau}_z+\cos\theta\cos\alpha\,\hat{\tau}_x-\cos\theta\sin\alpha\,\hat{\tau}_y].
\end{equation}
According
to Ref. \onlinecite{Bergeret2}, the DC-current is calculated
by i) modifying the Hamiltonian by adding to it a term proportional to
$\hat{I}_A$ and defining a generating function for the time-averaged
current; ii) going to Floquet space and calculating the long-time
Floquet evolution operator at times $nT (n>>1)$. This defines a
generalized Josephson energy and the averaged current is obtained by a
double derivative:
\begin{equation}
I_{DC}=\pdv{E(\eta,\mu)}{\mu}\Big|_{\eta,\mu=0}\pdv{E(\eta,\mu)}{\eta}\Big|_{\eta,\mu=0}
,
\end{equation}
with $\pm E(\eta,\mu)$ the eigenvalues of the matrix $\hat{M}=\frac{1}{T}\int_0^T\hat{H}_n(t)dt+\mu\hat{\tau}_z$. This is the the first term in the expansion of the effective evolution operator in the detuning from the n-th order resonance, expressed by the small parameter $(E_A-n\frac{\Omega}{2})/\Omega$. The Hamiltonian $\hat{H}_n(t)$ is such that:
\begin{equation}
\label{Hn}
\hat{H}_n(t)=e^{in\Omega t\hat{\tau}_z/2}[\hat{H}_A(t)+\eta \hat{I}_A-n\frac{\Omega}{2}\hat{\tau}_z]e^{-in\Omega t\hat{\tau}_z/2}
.
\end{equation}

After some simplifications, one obtains:

\begin{widetext}
\begin{equation}
\frac{1}{T}\int_0^Tdt\hat{H}_n(t)=\frac{1}{T}\int_0^Tdt\left[\hat{\tau}_z\left(E_A-n\frac{\Omega}{2}\right)
-\hat{\tau}_x\left(\frac{\dot{\theta}}{2\cos \alpha}+\eta \cos\theta\sin\alpha\pdv{|\Gamma|}{\varphi}\right)\sin(n\Omega t)\right]
\label{hamiltonian_parity}
.
\end{equation}
\end{widetext}

Let us first consider $\varphi=0$, with $b$ small. After several manipulations (see Appendix A), Eq. \eqref{hamiltonian_parity}
yields the effective RWA Hamiltonian close to the first-order resonance:
\begin{eqnarray}
\label{Effective_H_0}
\frac{1}{T}\int_0^T\hat{H}_1(t)dt&=&\hat{\tau}_z\left(\bar{E}_0-\frac{\Omega}{2}\right)\\
\nonumber
&+&\hat{\tau}_x\frac{\gamma_0 b }{2E_0}\Big[\Omega \cos\frac{\chi}{2}F_1-\eta\epsilon_0\sin\frac{\chi}{2}F_2\Big],
\end{eqnarray}
where $E_0=\sqrt{\epsilon_0^2+4\gamma_0^2}$ is the Andreev energy without microwaves, 
$\bar{E}_0=E_0(1-\frac{\gamma_0^2}{E_0^2}b^2\sin^2\frac{\chi}{2})$
and $F_1, F_2$ are constants defined in Appendix A. The eigenvalues of $\hat{M_1}$ verify: 
\begin{eqnarray}
\nonumber
E^2(\eta,\mu)&=&\left(\bar{E}_0-\frac{\Omega}{2}+\mu\right)^2\\
&+&\frac{\gamma_0^2 b^2 }{4 E_0^2}\Big[\Omega\cos\frac{\chi}{2}F_1-\eta\epsilon_0\sin\frac{\chi}{2}F_2\Big]^2
.
\end{eqnarray}
Finally,  the chiral current close to the first resonance is:
\begin{equation}
I_{chir,0}(\chi,\varepsilon_0)=\frac{e\gamma_0^2b^2\varepsilon_0}{E_0}\frac{(\Omega-2\bar{E}_0) F_1 F_2\sin\chi}{(\Omega-2\bar{E}_0)^2+4\gamma_0^2b^2F_1^2\cos^2\frac{\chi}{2}}
.
\end{equation}

This resonance is plotted in Figure \ref{fit}, together with the full Keldysh result
(see next Section). The resonance
width is given by $2\gamma_0bF_1|\cos\frac{\chi}{2}|$ and the maximal
chiral current is anharmonic in $\chi$:
\begin{equation}
\label{Imax}
I_{chir,0}^{max}\sim\frac{e\gamma_0bF_2|\varepsilon_0\sin\frac{\chi}{2}|}{2\hbar E_0}
\end{equation}

Let us consider now the case $\varphi=\pi$. The first harmonics $(\Omega =
2E_A)$ was calculated above to first order in the resonance
detuning. Eq.~(\ref{hamiltonian_parity}) is used and
$E_0=\left|\epsilon_0\right|$ stands for the Andreev state energy
without microwaves. Defining
$\bar{E}'_0=E_0(1+\frac{\gamma_0^2}{E_0^2}b^2\sin^2\frac{\chi}{2})$
leads to
\begin{eqnarray}
E^2(\eta,\mu)&=&\left(\bar{E}'_0-\frac{\Omega}{2}+\mu\right)^2\\\nonumber
&+&\frac{\gamma_0^2 b^2}{\pi^2 E_0}\left[\Omega\sin\frac{\chi}{2}G_1+\frac{2}{3}\eta \epsilon_0\cos\frac{\chi}{2}G_2\right]^2
\end{eqnarray}
and
\begin{eqnarray}
\nonumber
I_{chir,\pi}(\chi,\epsilon_0)=-\frac{8e\gamma_0^2 b^2\epsilon_0 }{ 3\pi^2 E_0 }\frac{(\Omega-2\bar{E}'_0)   G_1G_2\sin\chi}{(\Omega-2\bar{E}'_0)^2+\frac{16}{\pi^2}\gamma_0^2b^2G_1^2\sin^2\frac{\chi}{2}}
,\\
\end{eqnarray}
where $G_1, G_2$ are defined in Appendix A. 

 Comparing to $\varphi=0$, notice the sign change and the different $\chi$-dependence of the resonance width $\sim\frac{4}{\pi}\gamma_0bG_1|\sin\frac{\chi}{2}|$ and of the maximal chiral current $I_{chir,\pi}^{max}\sim \frac{2e\gamma_0bG_2}{3\pi\hbar}|\cos\frac{\chi}{2}|$. 
 
 Interestingly, the dependence in $\chi$ of $I_{max}$, Equation (\ref{Imax}), recalls that of an equilibrium junction made of a resonant dot, varying as $|\sin\frac{\varphi}{2}|$ due to closure of the Andreev gap at $\varphi=\pi$. In the present case, the ABS are gapped but at resonance the driven system behaves as gapless, generating the anharmonicity in $\chi$. 
 
 \subsection{Numerical solution}
\begin{figure}[ht]
\begin{center}
\includegraphics[width=19pc]{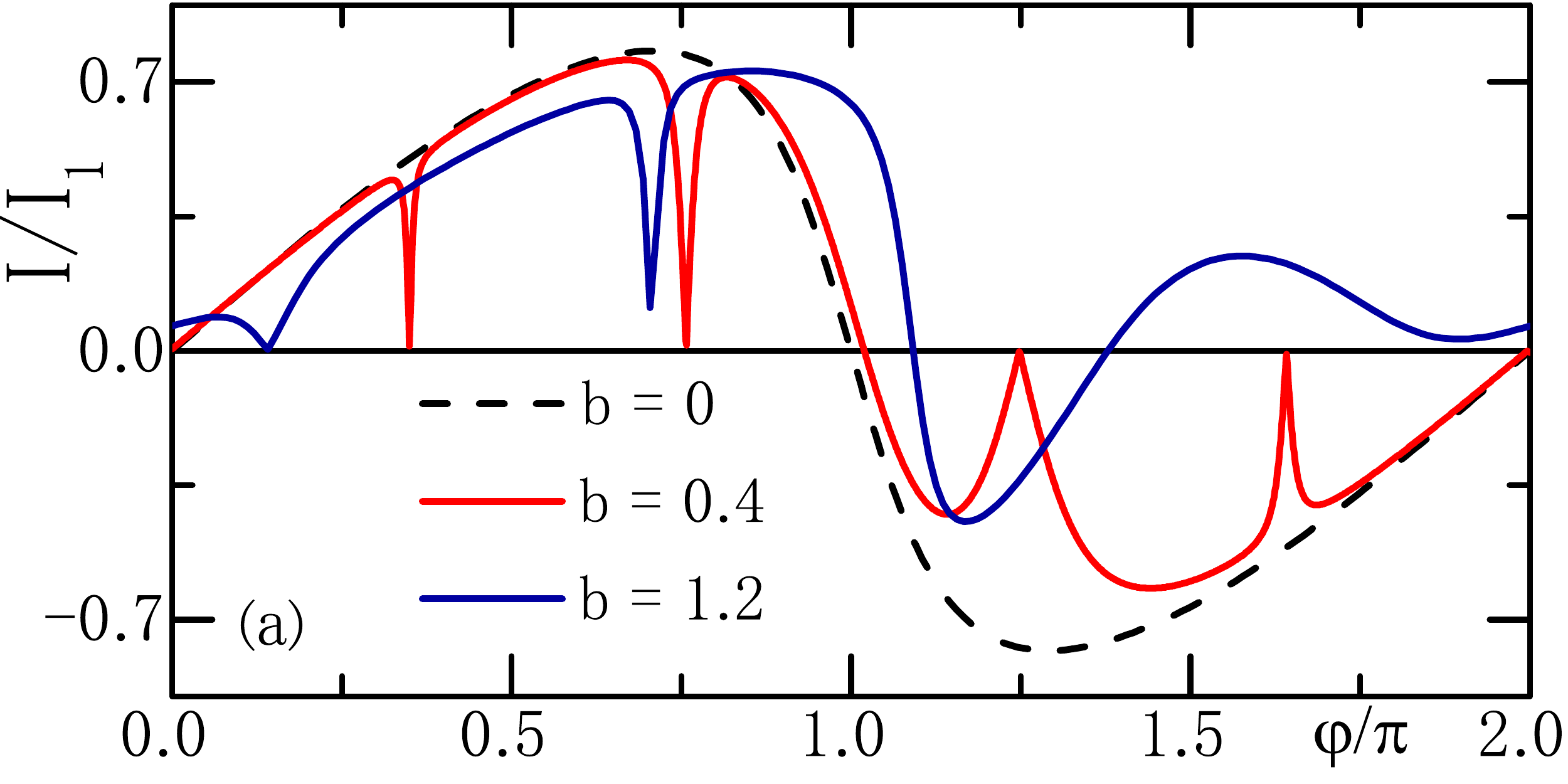} \\ 
\includegraphics[width=19pc]{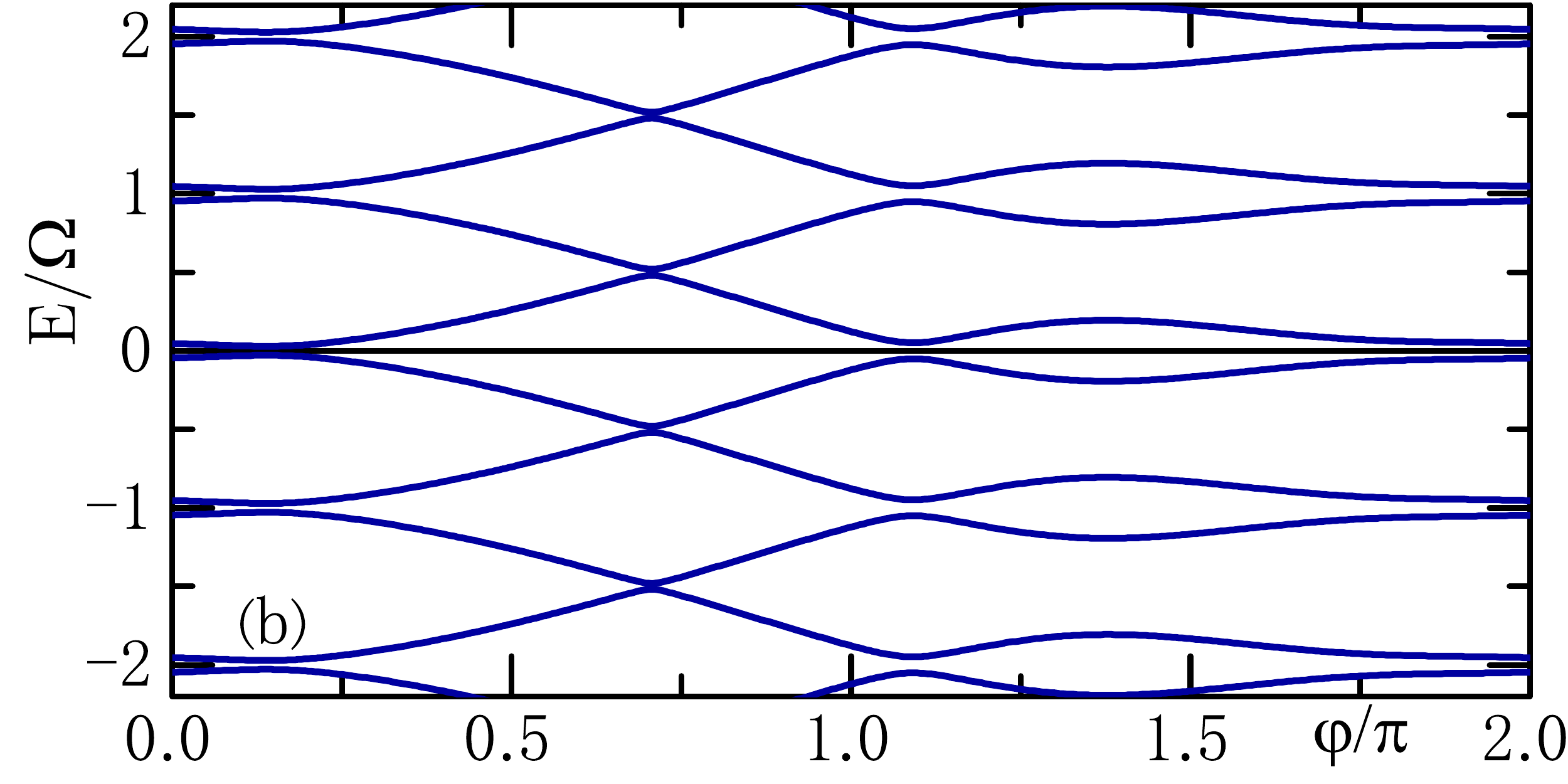} 
\caption{(Color online). (a) Current-phase relation in the chiral case ($\chi=\frac{\pi}{2}$), with $\varepsilon_0=0.5$, $\Omega=1.75$, $\gamma_1=\gamma_2=1$, $b_1=b_2=b=0, 0.4,1.2$, showing the nonadiabatic resonances and the chiral asymmetry around $\varphi=\pi$. (b) Floquet spectrum in the energy-phase plane, with $b_1=b_2=b=1.2$. The dotted lines indicate the equilibrium ABS.}
\label{Floquet_spectra}
\end{center}
\end{figure}

 As an example of a nonperturbative result, Fig. \ref{Floquet_spectra}a shows the Floquet current for different microwave amplitudes and Fig. \ref{Floquet_spectra}b shows the Floquet spectrum for $\chi=\frac{\pi}{2}$ and large $b_1=b_2$. We note $I_1 = \frac{e}{\hbar}\gamma_1$. The anticrossings are shifted in phase and clearly asymmetric, which reflects time symmetry breaking and the chirality. To calculate the time-averaged Josephson current, one needs in principle to fix initial conditions e.g.  perform an average over an initial distribution of Floquet eigenstates. Here we use a simple protocol, which perfectly maps onto the ground state Andreev current for zero microwave amplitude $b_{1,2}=0$. As $\varphi$ increases, the zero-th order Floquet state $\Psi_{n=0,-}$ is followed everywhere except in the centre of the anticrossing where it jumps across the anticrossing gap. As in Refs. \cite{Bergeret1,Bergeret2}, one obtains dips in the current when the ABS splitting is a multiple of the microwave frequency. Again, there is a strong asymmetry in the current, due to the chiral phase $\chi$. Most importantly, nonzero ``chiral'' currents are obtained $\varphi=0,\pi$.
 
\section{General solution: Keldysh analysis}
The microwave perturbs the ABS, also causing transitions towards the quasiparticle continuum, an effect negelcted in the infinite gap approximation. Let us fully solve the model Hamiltonian (\ref{Hamilt}) by using
 Keldysh nonequilibrium Green's functions. Those allow to obtain the
    spectral density and the dc current from
  Hamiltonian~(\ref{Hamilt})\cite{Nozieres,Cuevas,Regis1}. The
    Keldysh Green's function (GF) $\hat{G}^{+,-}$ is defined
    in the Nambu space spanned by the Pauli matrix
    $\hat{\tau}$ by:
    
   \begin{equation}
   \hat{G}_{ab}^{+-}(\tau,\tau')=i
   \begin{pmatrix}
  \langle
    c^{\dagger}_{b\uparrow}(\tau')c_{a\uparrow}(\tau)&\langle
    c^{\dagger}_{b\uparrow}(\tau')c^{\dagger}_{a\downarrow}(\tau)\\
    \langle
    c_{b\downarrow}(\tau')c_{a\uparrow}(\tau)\rangle&\langle
    c_{b\downarrow}(\tau')c^{\dagger}_{a\downarrow}(\tau)
    \end{pmatrix},
    \end{equation}
    and $\hat{G}_{ab}^{A}(\tau,\tau')$ obtained by replacing all
  correlators $\langle A(\tau')B(\tau)\rangle$ by
  $\theta(\tau-\tau')\langle \{A(\tau'),B(\tau)\}\rangle$ ($a,b=1,2, d$). Due to time periodicity, double
  Fourier transform is performed as:
 \begin{equation}
 \hat{G}_{ab,nm}(\omega)=\int\int d\tau d\tau' e^{i\omega(\tau-\tau')} e^{i\Omega(n\tau-m\tau')}\hat{G}_{ab}(\tau,\tau')
 \end{equation}

The Dyson equation implies a product in frequency space and a convolution product in the indices $n$: 
\begin{eqnarray}
\label{Dyson}
\nonumber
\hat{G}^{R,A}&=&\hat{g}^{R,A}+\hat{g}^{R,A}\hat{\Sigma}^{R,A}\hat{G}^{R,A}\\
\hat{G}^{+-}&=&(\hat{I}+\hat{G}^{R}\hat{\Sigma}^{R})\hat{g}^{+-}(\hat{I}+\hat{\Sigma}^{A}\hat{G}^{A})
\end{eqnarray}
where $\Sigma_{jd}^{R,A}$ and the bare Green's functions are defined in Appendix B.

The dc current is calculated between lead $1$ and the dot (the trace is in Nambu space):
\begin{eqnarray}
\nonumber
I =\frac{e}{\hbar}\trace\Big\{\sigma_z \int d\omega\sum_n\tilde{\Sigma}_{d1,n}\hat{G}^{+-}_{1d,0n}(\omega)\\
 - \tilde{\Sigma}_{1d,n}\hat{G}^{+-}_{d1,0n}(\omega)\Big\}
\end{eqnarray}
where $\tilde{\Sigma}_{jd,n}$ is the $n$th harmonic of the periodic
self-energy $\Sigma_{jd}=\Sigma_{jd}^{R,A}$. Details are given in Appendix B.

Solving for the Dyson equation and taking the $n=m=0$ component yields
the DC Josephson current $I(\varphi, \chi,\varepsilon_0)$, which is a
function of i) the superconducting phase difference
$\varphi$, ii) the chiral (microwave) phase
difference $\chi$, iii) the microwave amplitudes
$b_{1,2}$ and frequency $\Omega$ (one takes $\hbar=1$ in all the
figures), and iv) the dot parameters $\varepsilon_0$
and $t_{1,2}$. The current is in units of
  $I_0=\frac{e}{\hbar}\Delta$. The values $\eta_s=10^{-3}\Delta,
  \eta_d=10^{-5}\Delta$ are used for the inelastic parameters and temperature is $T=0$, unless
  specified otherwise.

\begin{figure}[ht]
\begin{center}
\includegraphics[width=19pc]{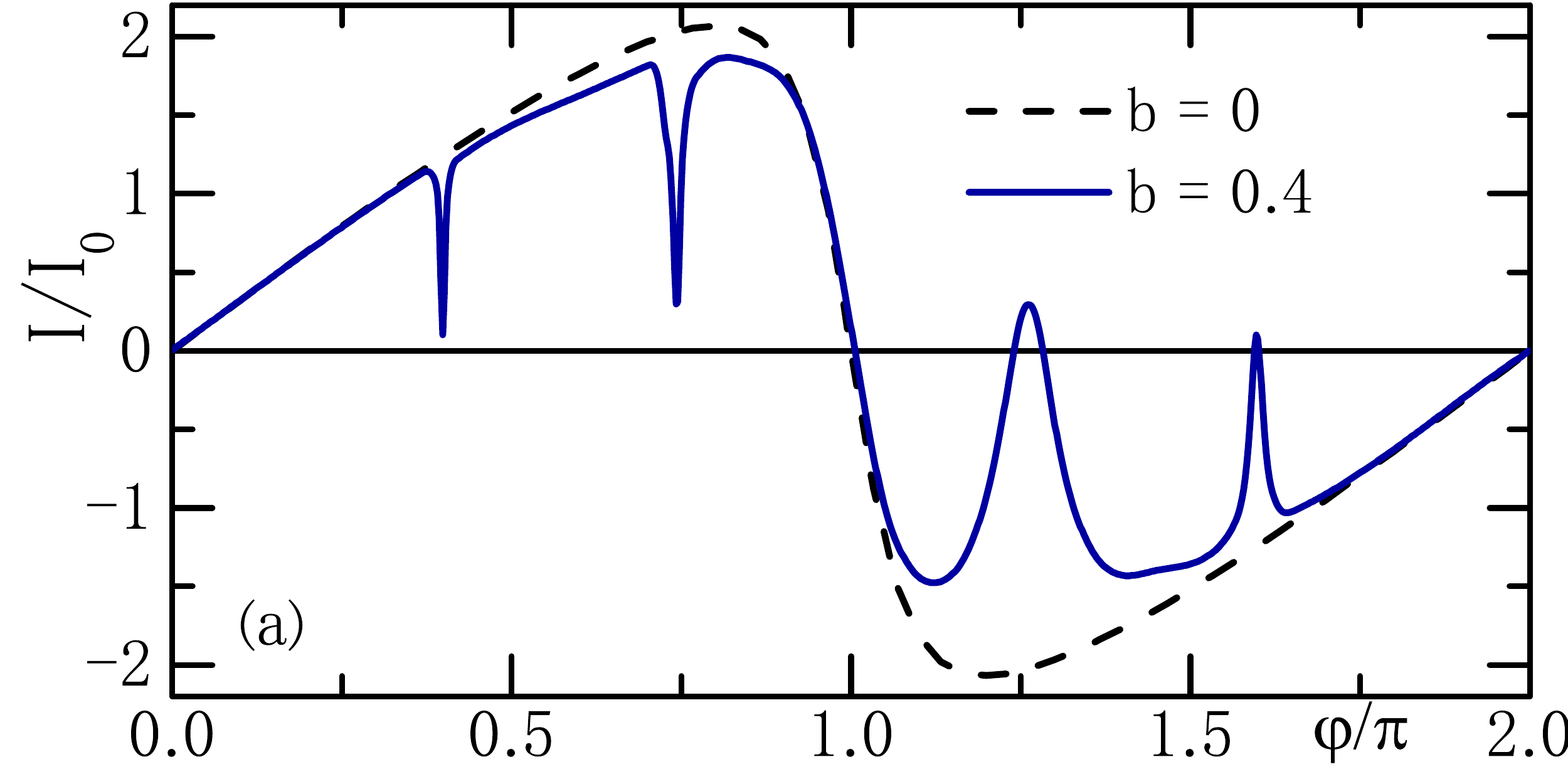} \\ 
\includegraphics[width=19pc]{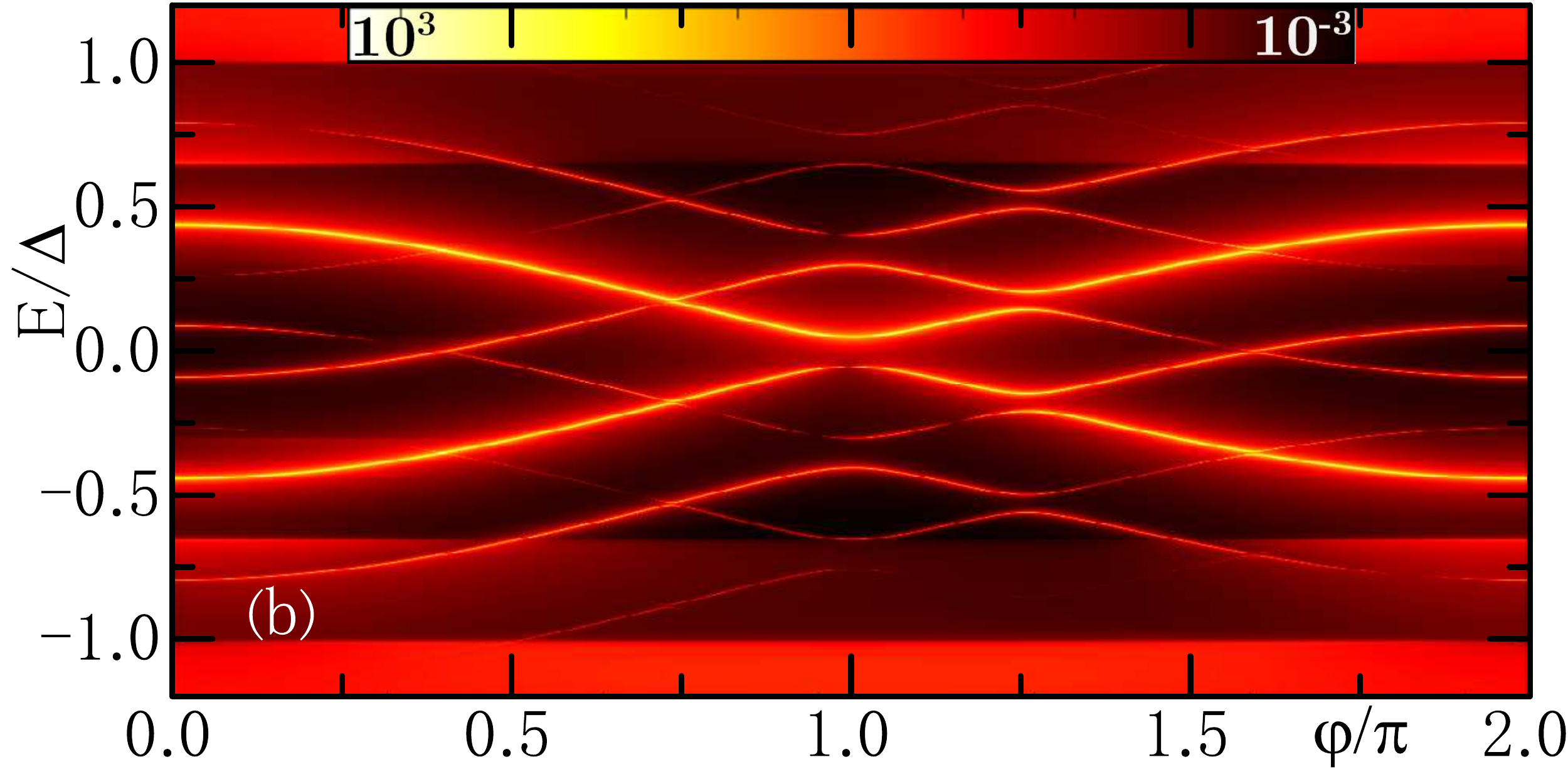} 
\caption{(Color online). (a) Current-phase relation in the chiral case ($\chi=\frac{\pi}{2}$), with $\varepsilon_0=0.1\Delta$, $\Omega=0.35\Delta$, $t_1=t_2=0.6\Delta$, $b_1=b_2=b=0.4$, showing the nonadiabatic resonances and the chiral asymmetry around $\varphi=\pi$. (b) Density of states (see text) in the energy-phase plane, with the same parameters.}
\label{current-phase}
\end{center}
\end{figure}

\begin{figure}[ht]
\begin{center}
\includegraphics[width=19pc]{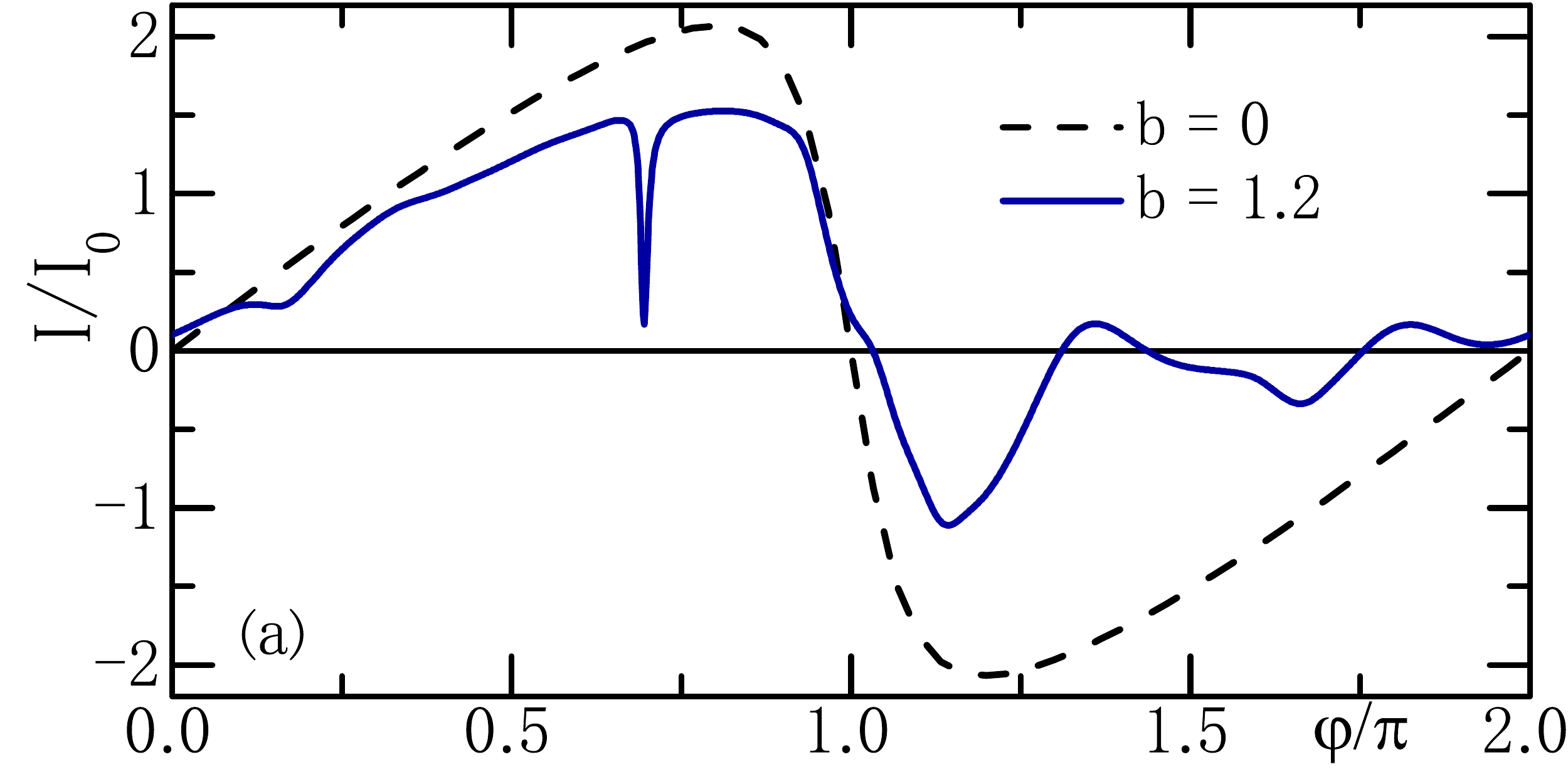} \\ 
\includegraphics[width=19pc]{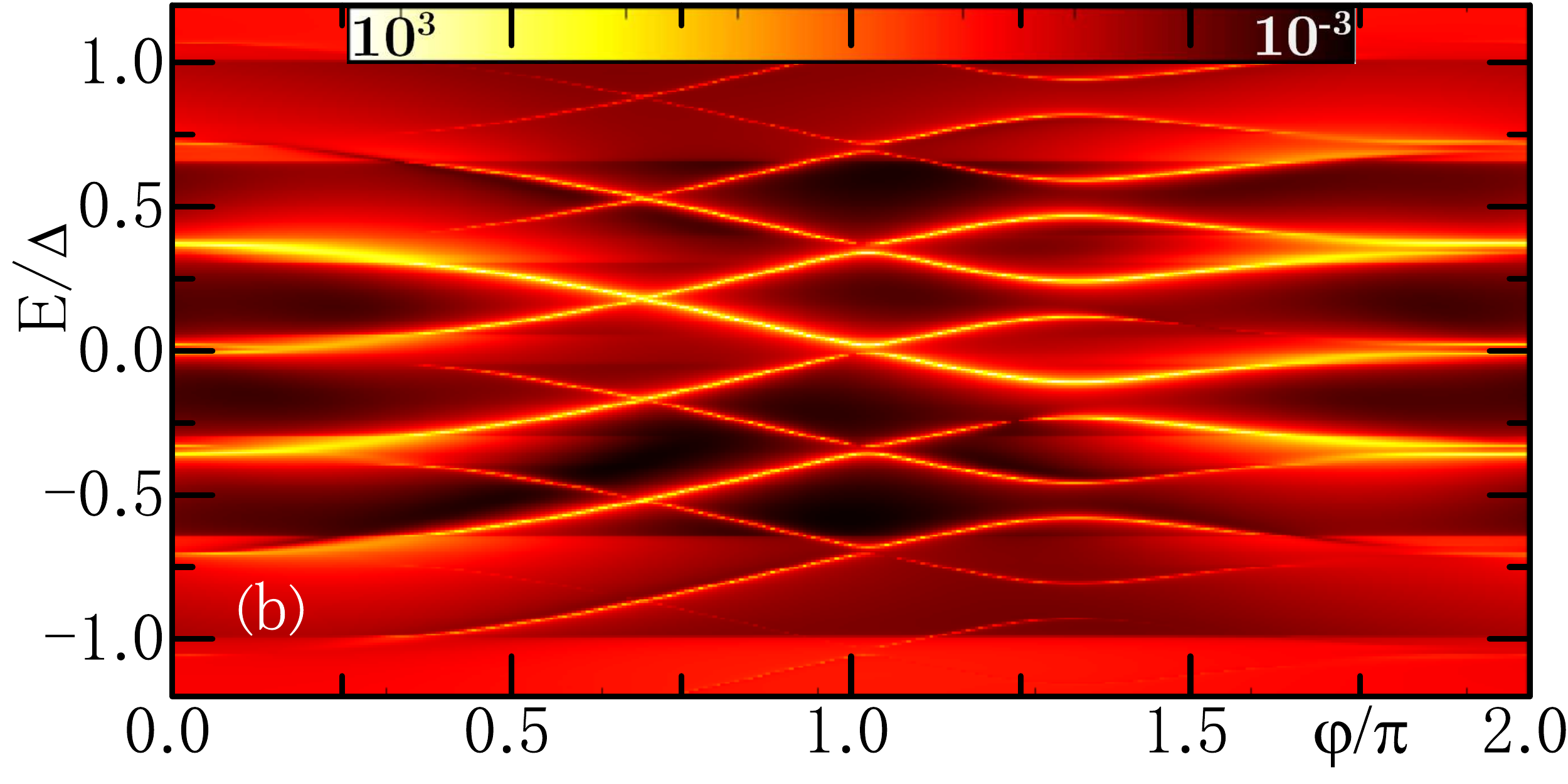} 
\caption{(Color online). Idem as Fig. (\ref{current-phase}), same parameters except $b_1=b_2=1.2$. The chiral currents at $\varphi=0,\pi$ are apparent despite the nonresonant behaviour.}
\label{current-phase2}
\end{center}
\end{figure}

Fig. \ref{current-phase}a shows the current-phase relation in the
chiral case, for moderate $b_1=b_2=0.4$. The evidence for chirality is
confirmed by plotting the effective density of states,
defined as $\rho_{d} = 2 \Im(G^{a}_{dd} -
G^{r}_{dd})$, where the anticrossings causing the resonances are asymmetric. This qualitatively confirms the trends obtained in the IGM (Section III, Fig. \ref{Floquet_spectra}). 
Notice the logarithmic scale in the amplitude. Some broadening is due to 
the inelastic parameters but it is mainly due to coupling to the continuum via the microwave excitation. This spectrum could be observed by microwave\cite{Bretheau} or
tunnel\cite{Pillet} ABS spectroscopy. The phase shift and the chiral currents (at $\varphi=0,\pi$) are very small for those excitation amplitudes. Fig. \ref{current-phase2} instead shows the case of a higher microwave amplitude. The phase shift and nonzero chiral currents are quite visible. Again, these features are similar to those found in the infinite-gap model (Fig. \ref{Floquet_spectra}). Since the microwave radiation couples strongly the equilibrium ABS to the continuum, only a qualitative agreement can be found.  

\begin{figure}[ht]
\begin{center}
\includegraphics[width=10pc]{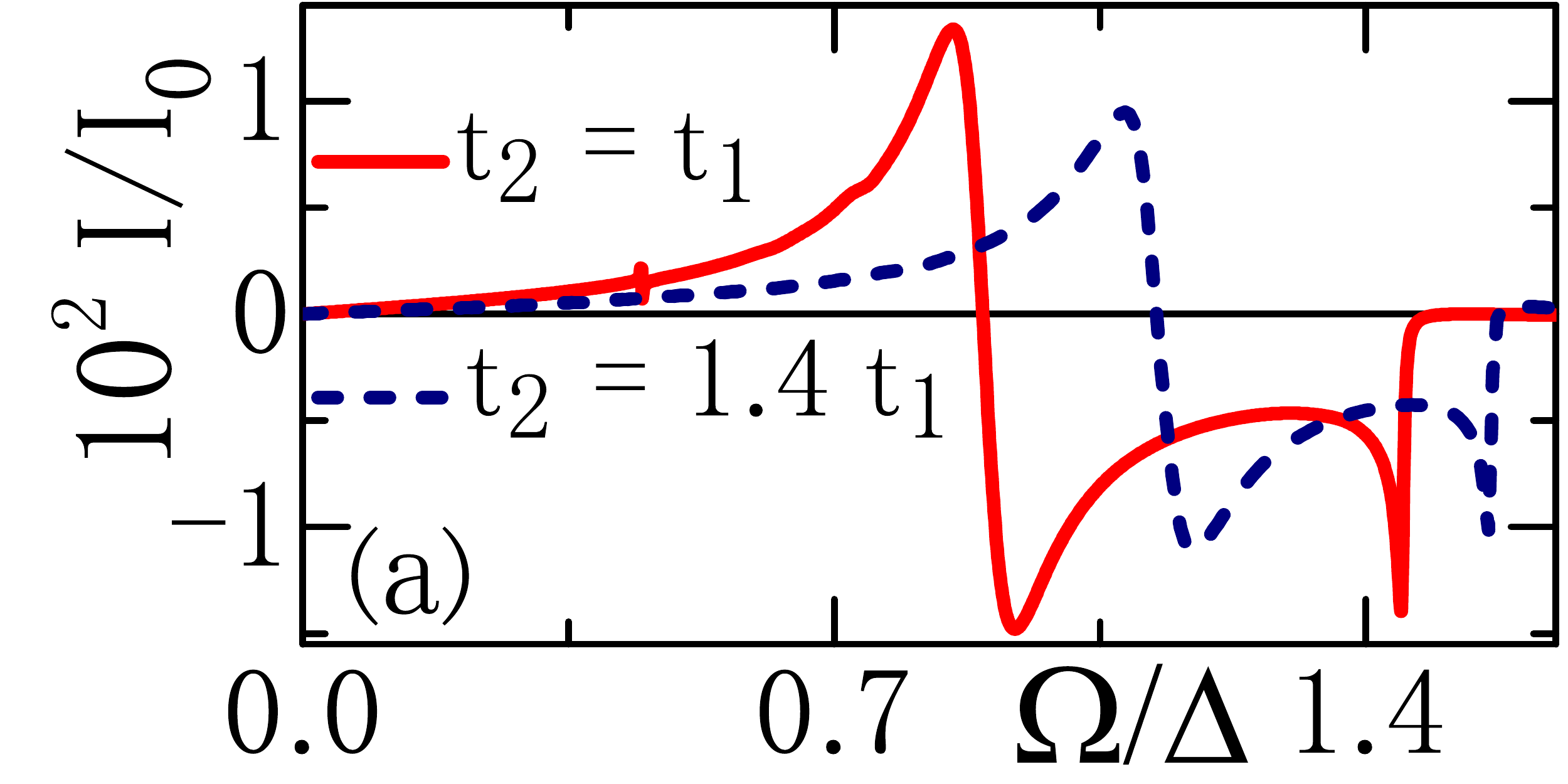}
\includegraphics[width=10pc]{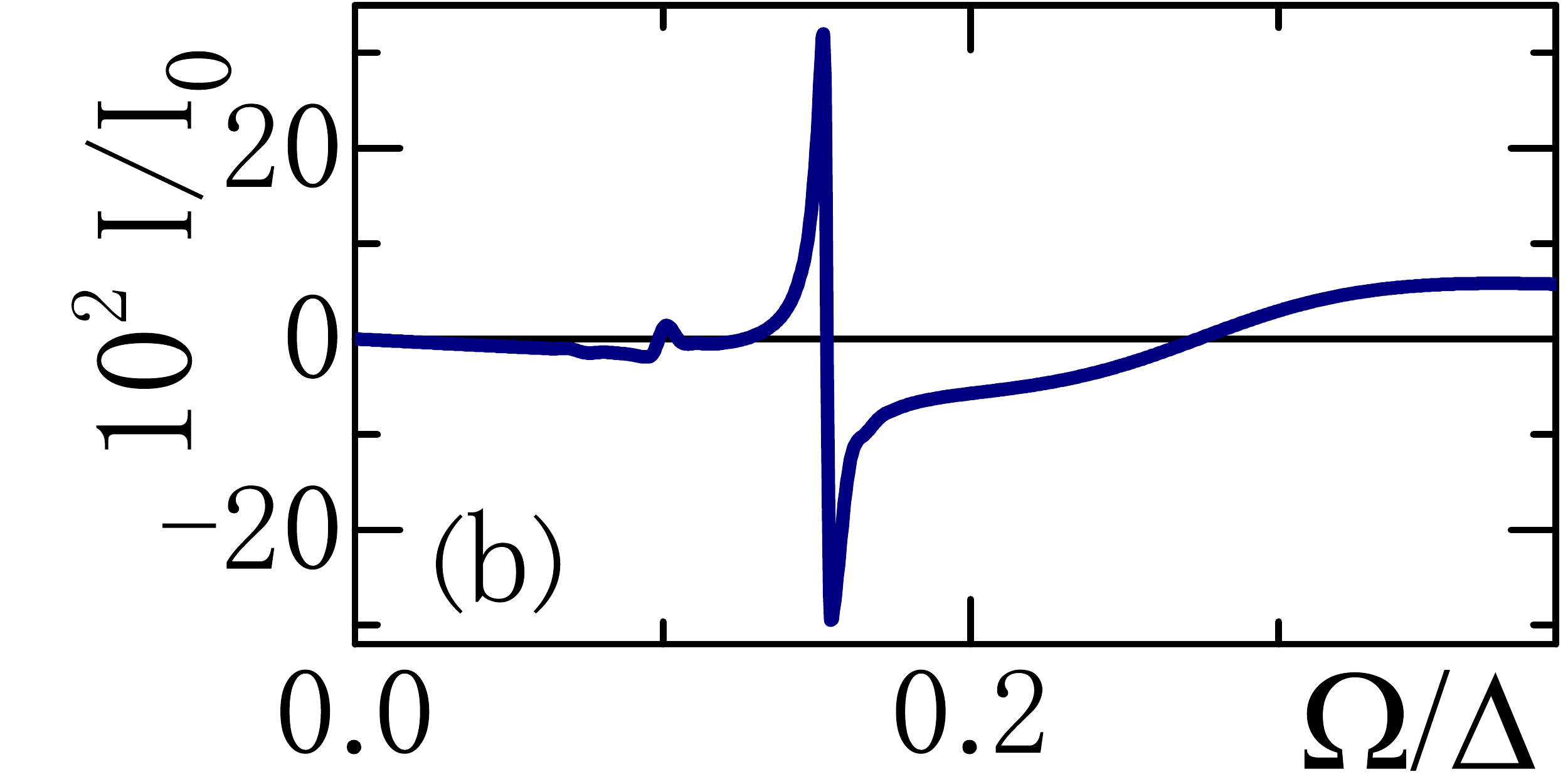}
\\[5px]
\includegraphics[width=10pc]{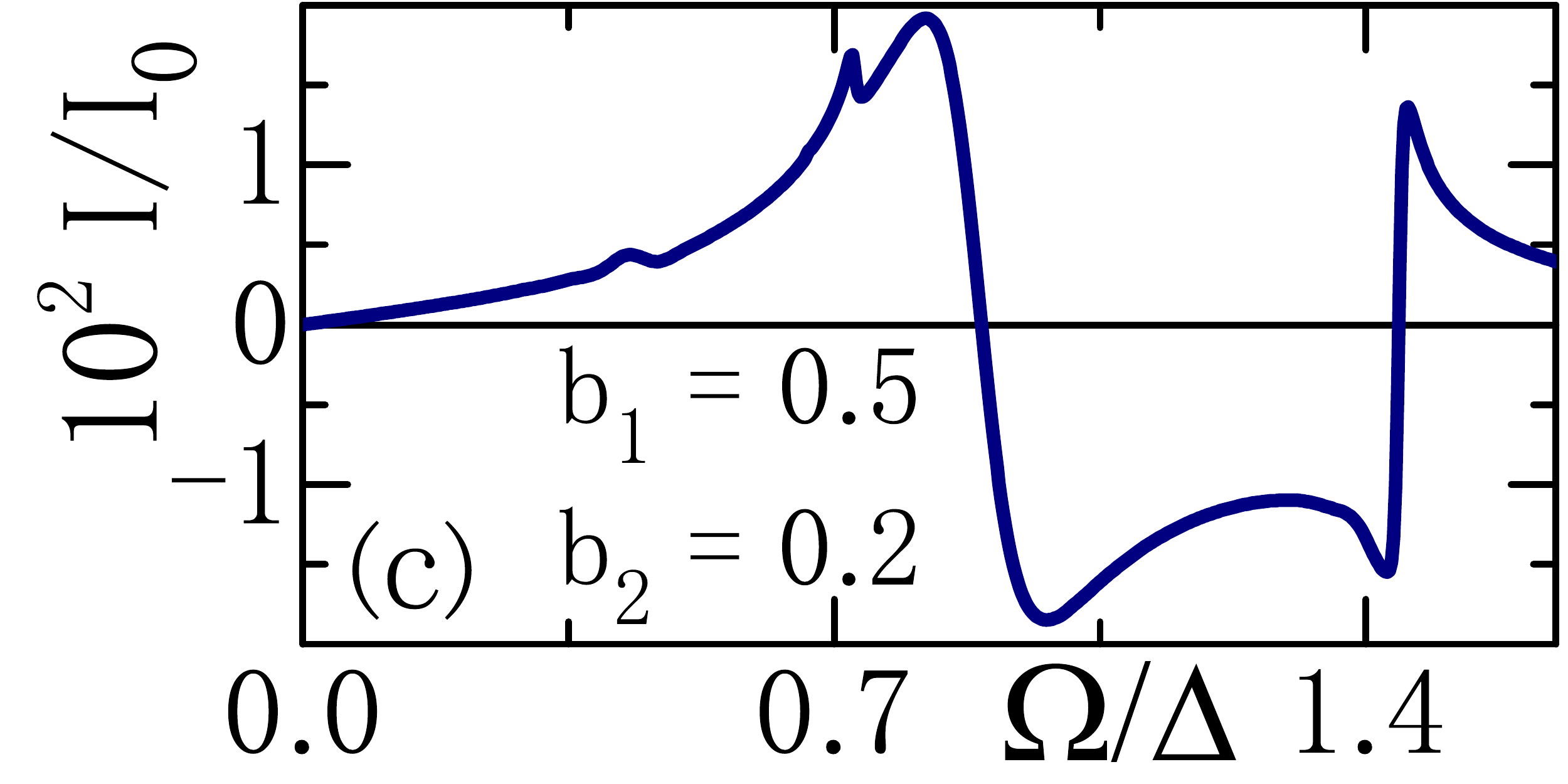}
\includegraphics[width=10pc]{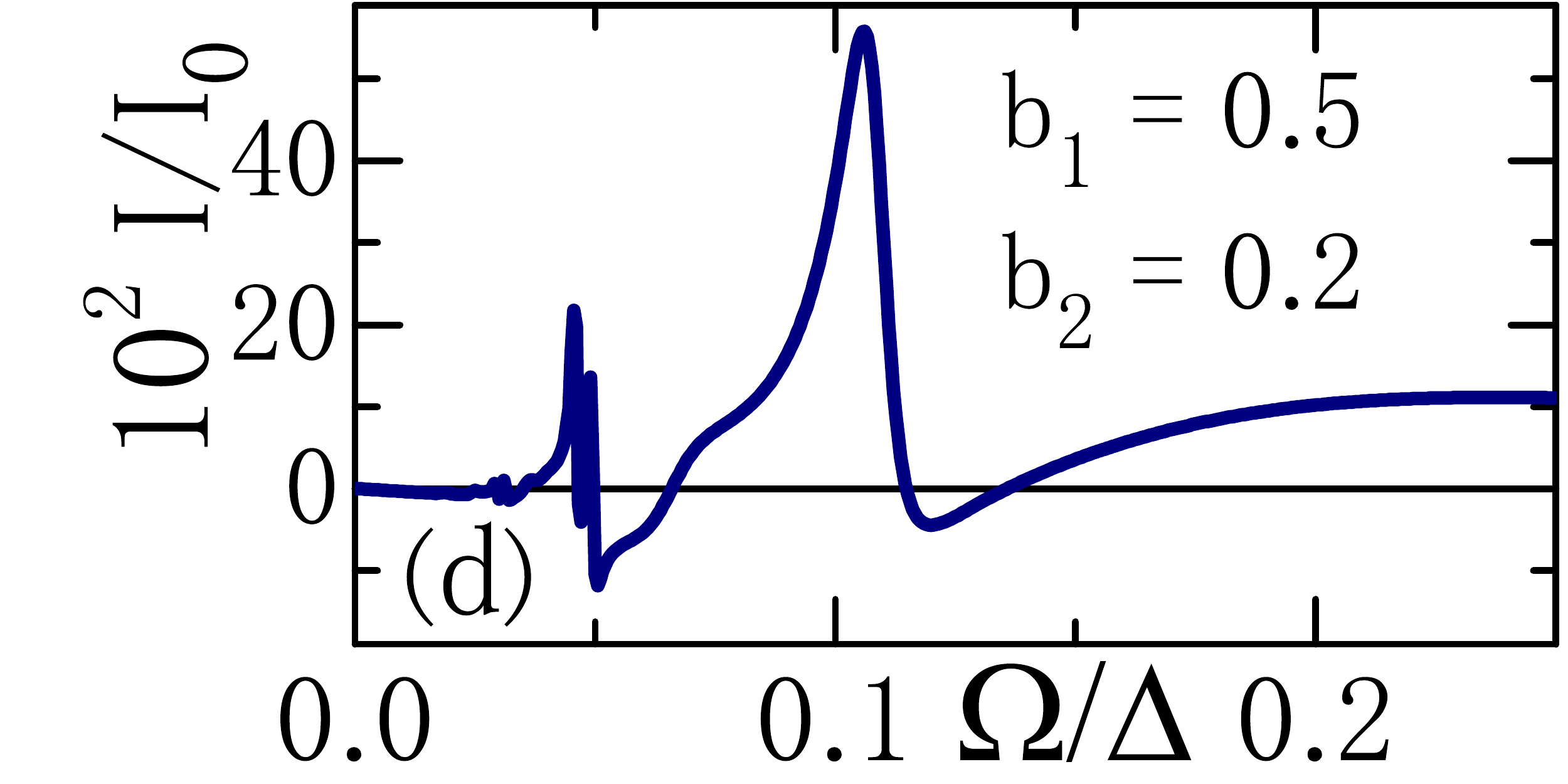}
\caption{(Color online). (a) Variation with $\Omega$ of the chiral current ($\varphi=0$, $\chi=\frac{\pi}{2}$), with $\varepsilon_0=0.1\Delta$, $t_1=0.6\Delta$, $b_1=b_2=0.2$: broad resonance around $\Omega=0.9\Delta$ (first harmonic), narrow asymmetric resonance due to the gap edge around $\Omega=1.4\Delta$. (b) Same but $\varphi=\pi$, $\varepsilon_0=0.4\Delta$, $t_1=t_2=\Delta$: resonances from right to left: broad (1st harmonic, $\Omega\simeq0.3\Delta$), sharp (2nd harmonic, $\Omega\simeq0.15\Delta$), small (3rd harmonic, $\Omega\simeq0.1\Delta$). (c) Same parameters as (a) except $b_1=0.5,b_2=0.2$. (d) Same as (b) except $b_1=0.5,b_2=0.2$.}
\label{current-frequency}
\end{center}
\end{figure}

\begin{figure}[ht]
\begin{center}
\includegraphics[width=10pc]{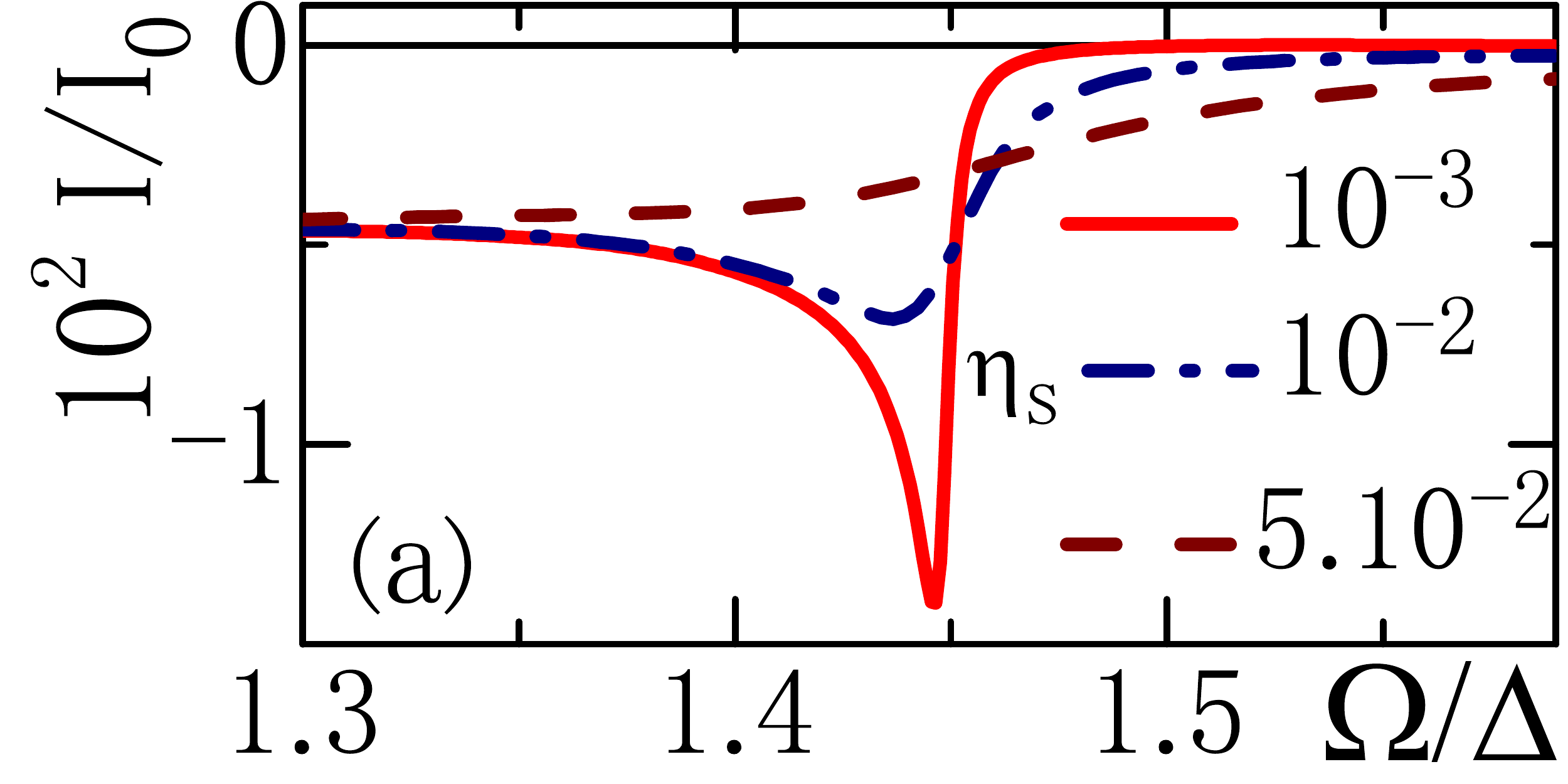}
\includegraphics[width=10pc]{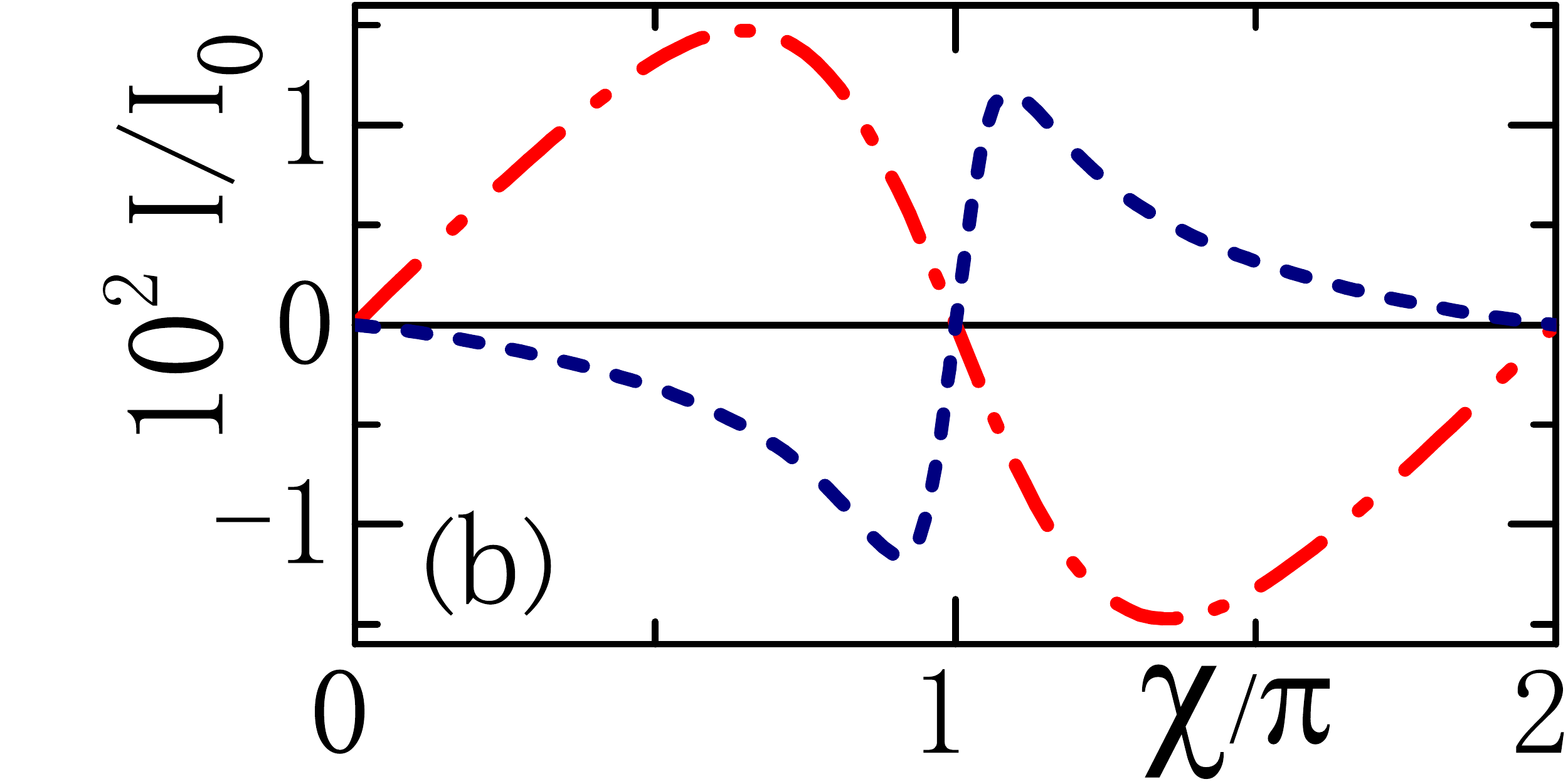}
\\[5px]
\includegraphics[width=10pc]{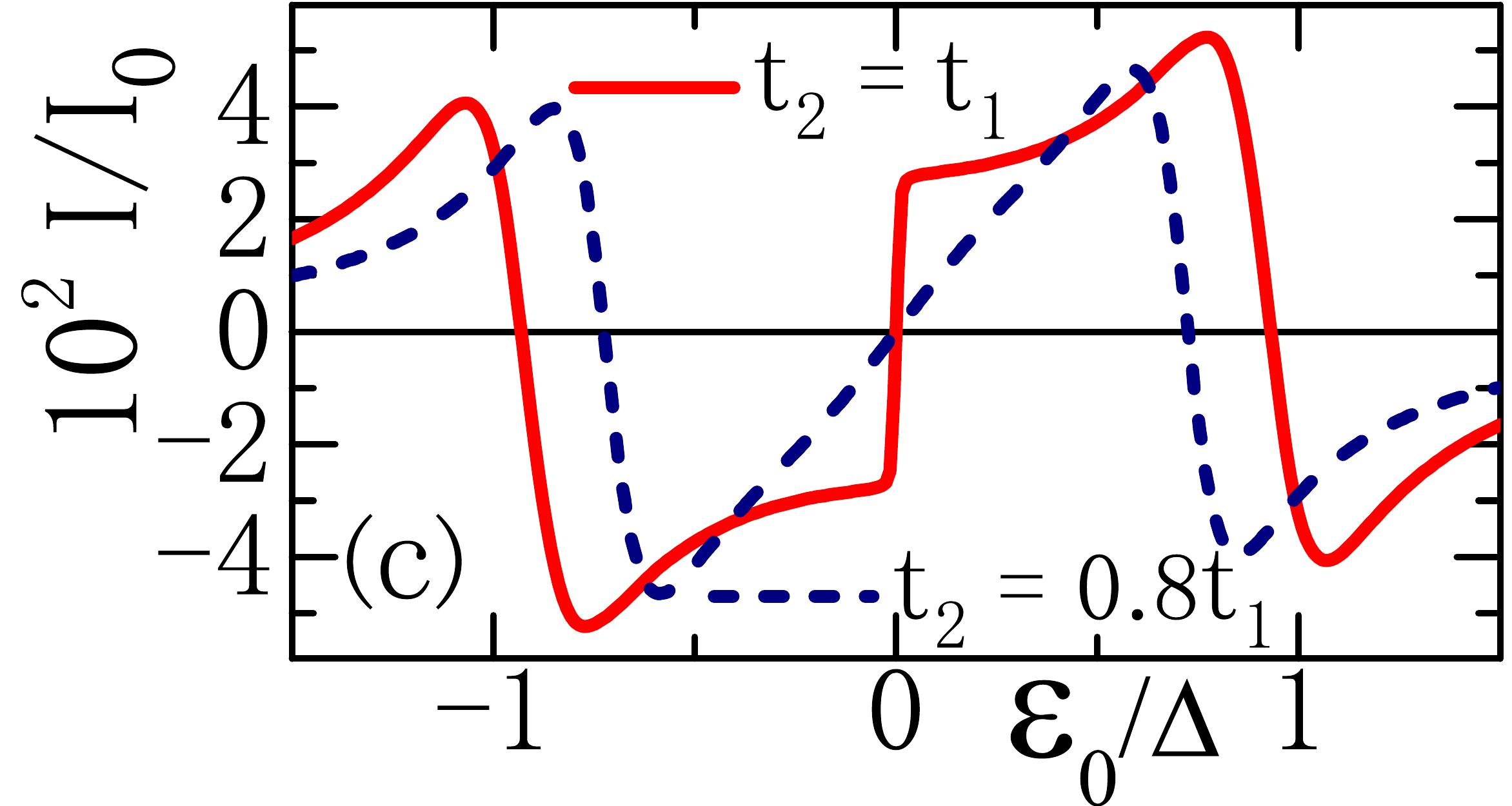}
\includegraphics[width=10pc]{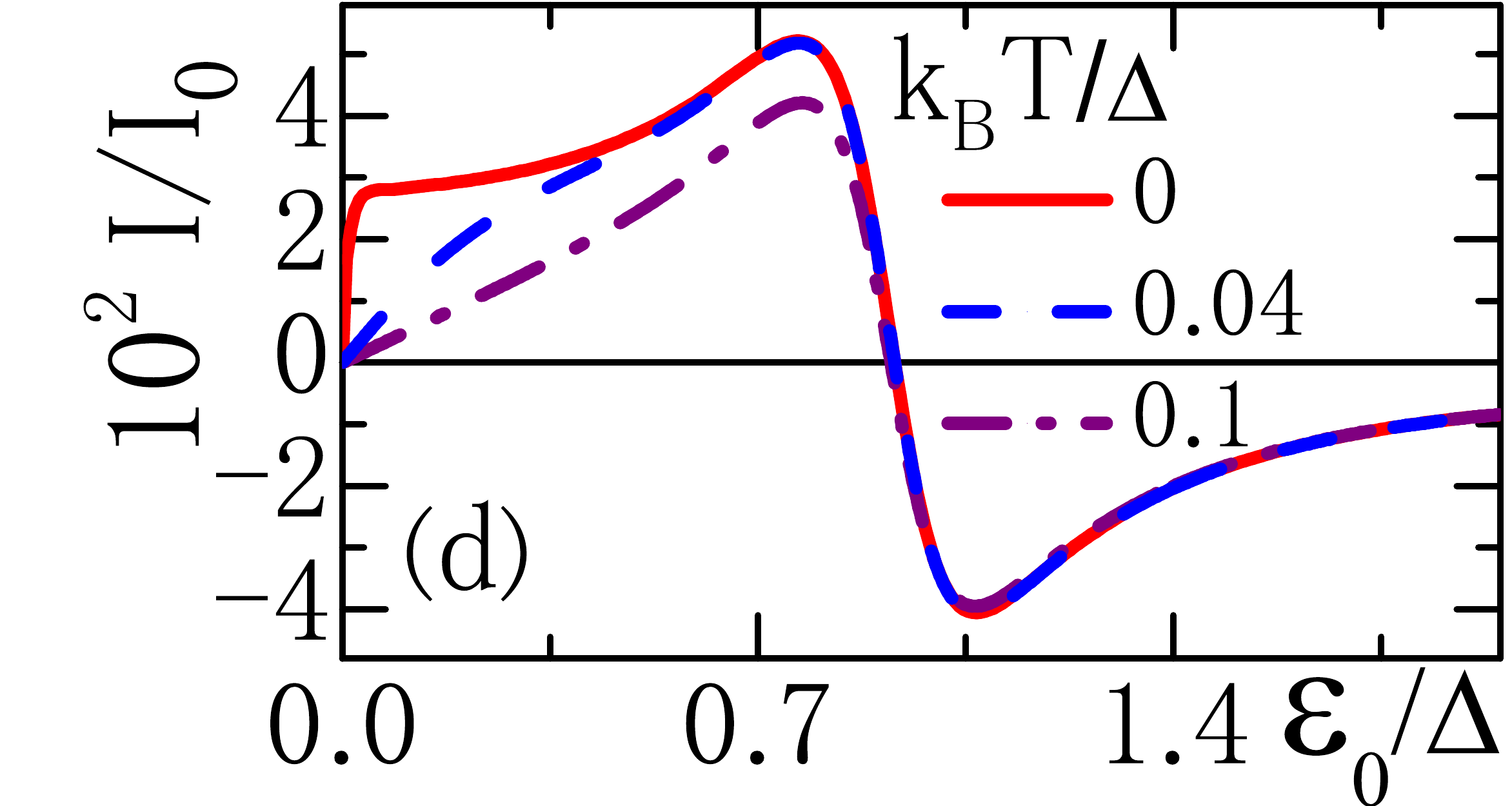}

\caption{(Color online). (a) $\eta_s$-broadening of the ABS-to-gap
  edge resonance at $\varphi =0$: same parameters as in
  Fig. \ref{current-frequency}a with $t_1=t_2$. (b) Dependence on
  $\chi$ ($\varphi=0$), with $\varepsilon_0=0.1\Delta$,
  $t_1=t_2=0.6\Delta$, $b_1=b_2=0.2$ and $\Omega=0.85\Delta$ (red
  dot-dashed, close to an extremum in $\Omega$) or $\Omega=0.9\Delta$
  (blue dashed, close to the resonance center). (c) Dependence on
  $\varepsilon_0$ ($\varphi=\pi,\chi=\frac{\pi}{2}$), with
  $\Omega=0.6\Delta$, $t_1=\Delta$, $t_2=t_1$ or $t_2=0.8t_1$,
  $b_1=b_2=0.2$. (d) Temperature dependence (same as (c), $t_2=t_1$).}
\label{current-dotlevel}
\end{center}
\end{figure}

Let us now set $\varphi=0$ or $\pi$ and analyze the chiral current $I_{chir}$. It is
amplified when a resonance occurs at $\varphi=0$ or
$\pi$. We focus here on small microwave amplitudes which lead to small chiral currents 
but display more clearly the main quantitative features. This situation can be compared to an equilibrium tunnel junction 
where the curent is harmonic except for a resonant dot. Fig. \ref{current-frequency} indeed
shows $I_{chir,0}(\Omega)$ and $I_{chir,\pi}(\Omega)$ as a
function of $\Omega$. The chiral current changes sign at resonance. For $\varphi=0$, the main resonance indeed occurs at
$\Omega\sim2E_{A}(\varphi=0)$, e.g. matching
the ABS spacing. A thin and asymmetric resonance also appears around $\Omega=1.4\Delta$, due to
a transition from the lowest ABS to the upper gap edge. Contrarily to the main resonance between ABS, 
it depends strongly
on the gap smearing parameter $\eta_s$ as shown in
Fig. \ref{current-dotlevel}a. The subtle dependence of a resonant
property with the coupling to quasiparticle states and with the
$\eta$'s has been discussed in
  Ref. \onlinecite{Regis2} for a related problem.  In the case
$\varphi=\pi$, the first harmonic resonance around $\Omega=0.3\Delta$ is quite soft and, remarkably,
an intense and narrow second harmonic ($\Omega\sim E_A(\varphi=\pi)\simeq0.15\Delta$)
resonance appears.

In the above-described pumping mechanism, the phase $\chi$ replaces $\varphi$ as the driving phase for the
Josephson current. Far from resonance, $I_{chir}$ is approximately
sinusoidal with $\chi$. Close to resonance, strong
nonharmonicity and a change of sign are instead obtained
(Fig. \ref{current-dotlevel}b). Fig. \ref{current-dotlevel}c shows the
chiral current as a function of the dot level $\varepsilon_0$ at fixed
$\varphi=\pi$ and $\chi=\frac{\pi}{2}$, displaying the symmetry
expressed by Eq. \eqref{sym2}. More generally, the
  symmetries \eqref{sym1} and \eqref{sym2} were numerically checked for $\varphi\neq 0$. Remarkably, the rapid change close to
$\varepsilon_0=0$ is in agreement with the analytical formula
(Equation (\ref{courant_pert})) obtained within the IGM.  Moreover, asymmetric
$\gamma$'s, or a nonzero temperature, makes $I_{chir}$ linear with
$\varepsilon_0$ (Fig. \ref{current-dotlevel}c). This behaviour reminds that of a resonant symmetric equilibrium junction 
where at zero temperature the current experiences a jump at phase $\varphi=\pi$. 

\begin{figure}
\includegraphics[width=19pc]{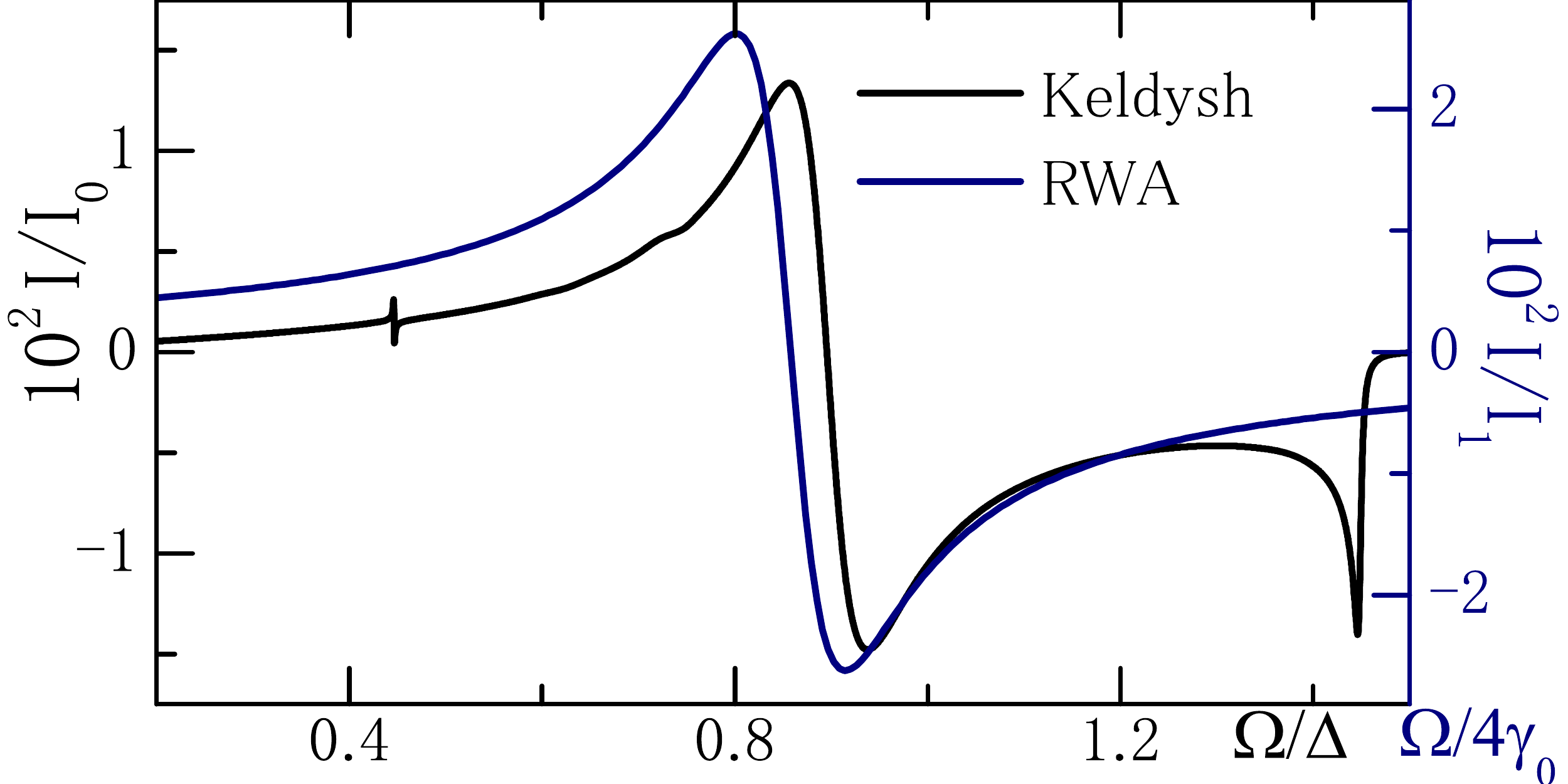}
\caption{(Color online) Chiral current at $\varphi=0$, $\chi=\frac{\pi}{2}$ close to the first harmonic resonance. Keldysh result ($\varepsilon_0=0.2\Delta$, $t_1=t_2=1.0\Delta$, $b_1=b_2=0.2$) vs RWA (infinite gap) with fitted parameters $\gamma_0=1.0$, $b=0.2$, $\epsilon_0=0.8$.}
\label{fit}
\end{figure}

Let us comment in more detail on the vicinity of a
  resonance such that $n\Omega\sim 2E_A(\varphi)$,
  with $\pm E_A(\varphi)$ the equilibrium Andreev bound state
energies. The salient result, featured in
Fig.~\ref{current-frequency}, is the maximal chiral
current close to the resonance, and
  its rapid sign change as the resonance is crossed. The exact calculation qualitatively agrees well with the RWA calculation in the IGM (Section IIIC). 
This is illustrated in Figure \ref{fit}. No quantitative agreement is possible due to the renormalizing effect of the quasiparticles. Yet, except for the resonance towards the continuum, one can nearly match the 
results of the Keldysh calculation with the IGM by fitting the parameters of the latter. 

The resonant chiral pumping effect found above is
robust against nonzero temperature ($k_BT=0.1\Delta$ here) and
junction asymmetry (Fig. \ref{current-dotlevel}c), and also nonsymmetric microwave amplitudes (Fig. \ref{current-frequency}c, d). 
It bears some
similarity with pumping mechanisms, as explored in a variety of
situations with normal or superconducting islands\cite{resonant_pumping}. Yet, it is
remarkable that the chiral current
emerges only beyond the adiabatic regime
and is maximal if the microwave
frequency -- or its harmonics --
  matches the ABS spacing, which is precisely an antiadiabatic
effect. For small $\varepsilon_0$, the resonant chiral current is
larger at $\varphi=\pi$ than at $\varphi=0$, due to the nonharmonicity
of the equilibrium $I(\varphi)$ close to $\varphi=\pi$.

\begin{figure}[ht]
\begin{center}
\includegraphics[width=10pc]{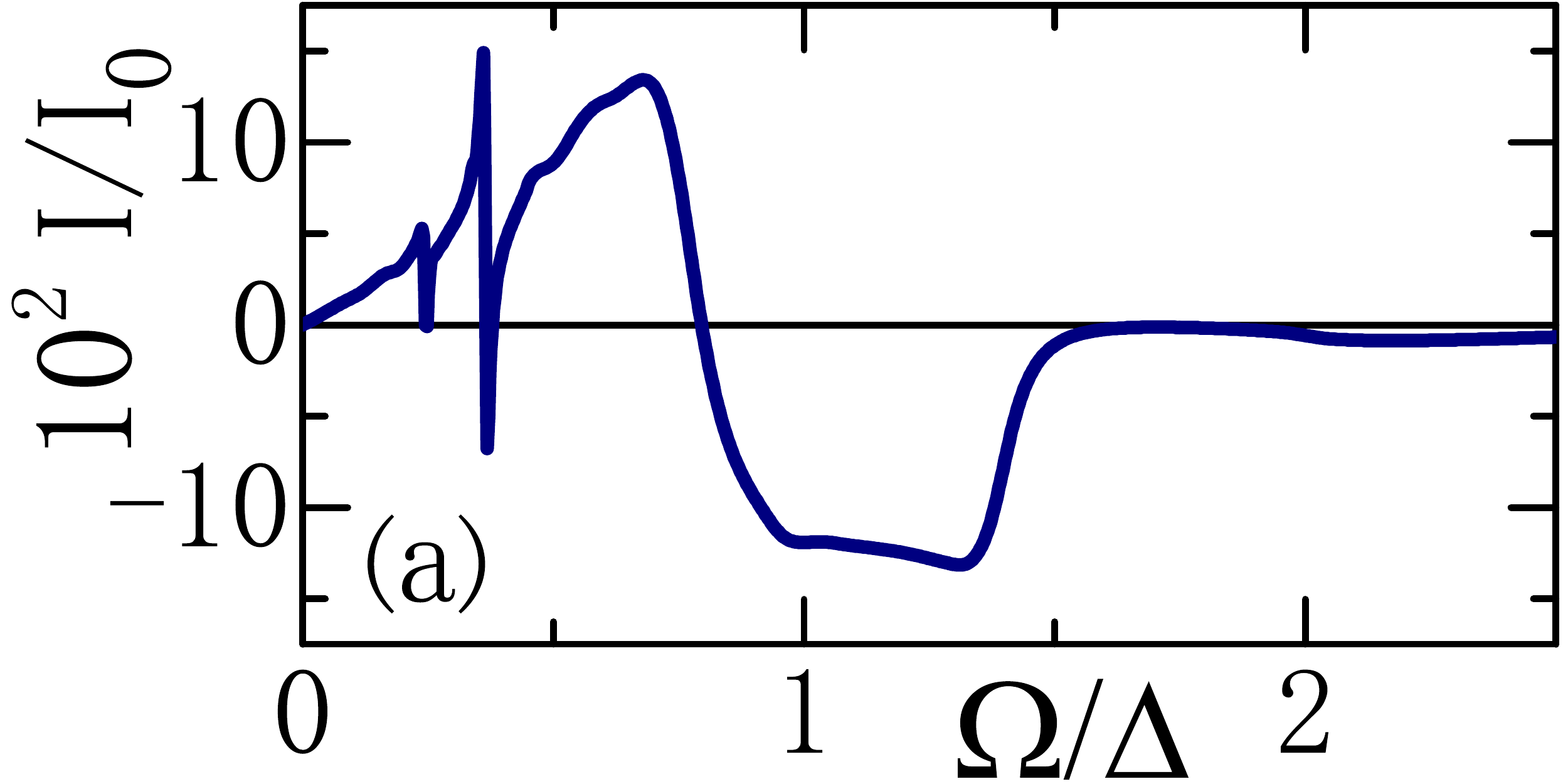}
\includegraphics[width=10pc]{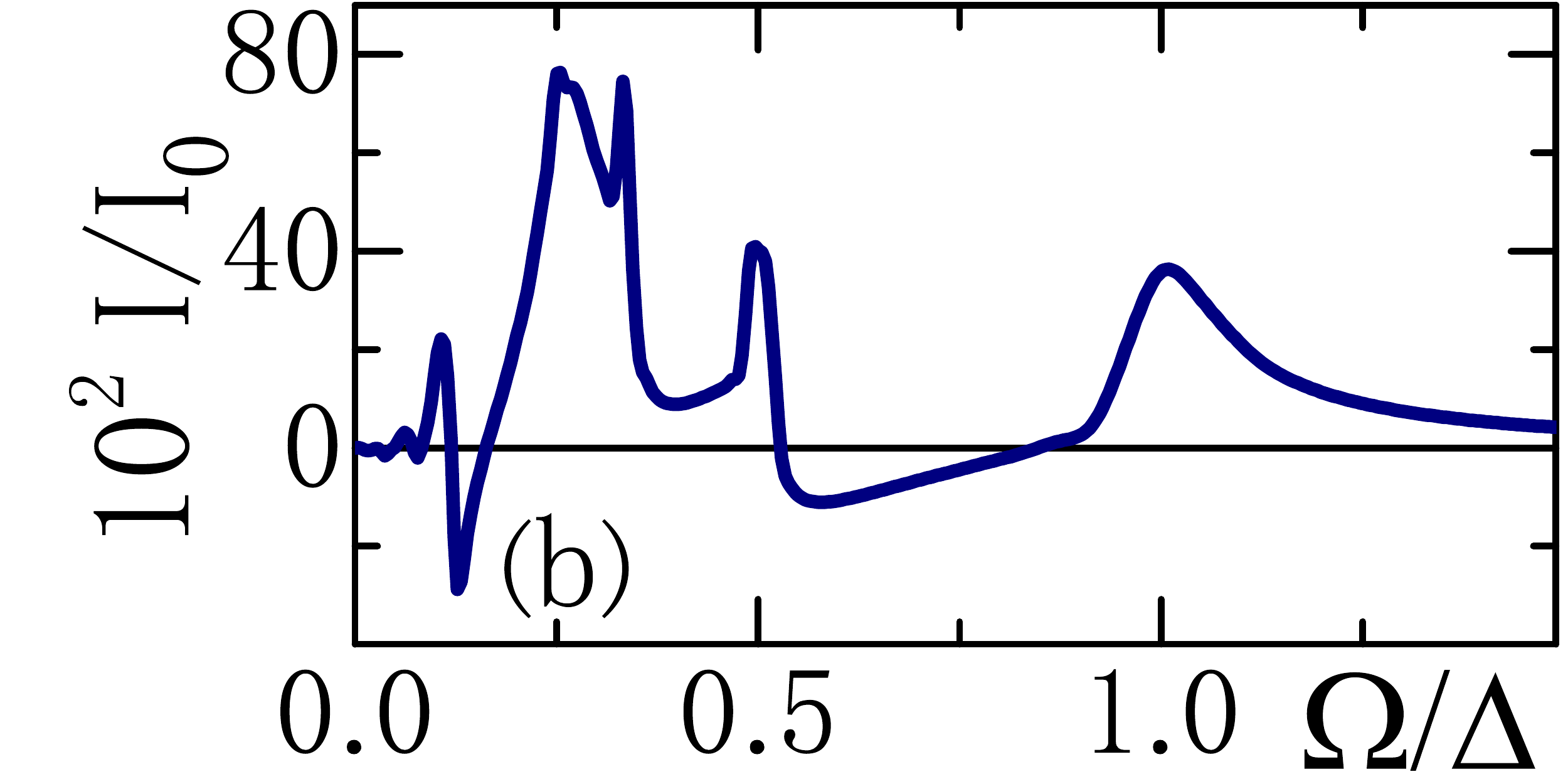}
\caption{(Color online). (a) Variation with $\Omega$ of the chiral current ($\varphi=0$, $\chi=\frac{\pi}{2}$), with $\varepsilon_0=0.1\Delta$, $t_1=t_2=0.6\Delta$, $b_1=b_2=1.2$ (b) Same but $\varphi=\pi$.}
\label{strongb}
\end{center}
\end{figure}

Fig. \ref{strongb} shows the frequency dependence of the chiral current for a stronger microwave amplitude ($b_1=b_2=1.2$). The resonances are much broader, and above all, very sizeable chiral currents are obtained, close to $0.8 I_0$ for $\varphi=\pi$. 

The $\chi$-dependence of the chiral current
can be compared to the $\varphi$-dependence of the
junction equilibrium current (e.g. with no microwave). The microwave amplitude conspires with the junction 
transparency to control the current amplitude far from resonance. This is similar to an equilibrium junction with a non resonant dot. 
Indeed, the $\sin\chi$
variation resembles the $\sin\varphi$ variation obtained for a
nonresonant dot. On the other hand, at a maximum close to resonance,
the strongly nonharmonic $\sin\frac{\chi}{2}$ variation (for
$\varphi=0$) resembles that obtained for a symmetric resonant dot
junction if $\varepsilon_0=0$: there, closure of the
Andreev gap at $\varphi=\pi$ results in a sawtooth
jump of the equilibrium current at zero temperature, which is rounded
by asymmetry and temperature (Fig. \ref{current-dotlevel}). A similar situation is
  met in the chiral case, with a nonzero $\varepsilon_0$
but chiral microwave resonant
with the ABS spacing. While in the non-resonant case the $\chi$ variation goes like $\sin\chi$, in the resonant regime it approaches $|\sin\frac{\chi}{2}|$. 
This interpretation is confirmed by the
asymmetry and temperature rounding of the jump at $\varepsilon_0=0$
(Eq. \ref{courant_pert} and Fig. \ref{current-dotlevel}c, d). Moreover,
the change of sign in $I_{chir}(\chi)$
upon crossing resonance can be compared to the
change of an equilibrium $I(\varphi)$ from $0$ to $\pi$ character.

For stronger microwave amplitudes, the current-phase relation is strongly anharmonic and the resonances much broader\cite{Bergeret1,Bergeret2}, 
and the same is true for the chiral current. 

Also,
a chiral current persists in presence of cross-talk between the two
dephased microwave excitations and towards the dot gate.This can be
shown by setting, instead of Eq.~(\ref{eq:1}),
$\varphi_1=\frac{\varphi}{2}+b_1\cos(\Omega
t+\frac{\chi}{2})+b'_1\cos(\Omega t-\frac{\chi}{2})$,
$\varphi_2=-\frac{\varphi}{2}+b_2\cos(\Omega
t-\frac{\chi}{2})+b'_2\cos(\Omega t+\frac{\chi}{2})$,
$\varepsilon_0(t)=\varepsilon_0+\varepsilon_1\cos(\Omega
t+\frac{\chi}{2})+\varepsilon_2\cos(\Omega
t-\frac{\chi}{2})$. Generalizing the large-$\Omega$ calculation, one obtains (Appendix C):
\begin{equation}
\label{currentchirCT}
I_{chir}=\pm\frac{2e}{\hbar^2}\frac{\gamma_1\gamma_2(b_1b_2-b'_1b'_2)\varepsilon_0 \sin\chi}{\Omega E_0^{eff}}
\end{equation}
where $E_0^{eff}$ is an effective ABS energy depending on the dot level modulations $\varepsilon_{1,2}$. The chiral current is thus robust against small crosstalk $b'_{1,2}\neq0$.

\section{Lattice chain mapping}
The Cooper pair pumping mechanism, illustrated in the IGM, can be understood in relation with charge pumping in some tight-binding chain models. For this purpose, let us introduce the number state representation $|N,\nu\rangle$ where $N$ is the number of pairs exchanged through the junction from terminal $1$ to terminal $2$, and $\nu=0,1$ indicates the charge state of the dot. The variable $N$ is by convention defined here from the pair numbers $N_{1,2}$ by $N=\frac{N_2-N_1-\nu}{2}$. The variable $N$ is conjugated of the superconducting phase difference $\varphi$. As a result, it is straightforward to rewrite the Hamiltonian $H_{\infty}$ in the number basis as:

 \begin{widetext}
\begin{equation}
H_{\infty}
=\sum_N \varepsilon_0\big(|N,1\rangle\langle N,1|-|N,0\rangle\langle N,0|\big)
+\sum_N\big(\gamma_1e^{-i\varphi_1(t)}|N,1\rangle\langle N,0|+\gamma_2e^{-i\varphi_2(t)}|N-1,1\rangle\langle N,0|
+ H. c.\big)
\end{equation}
\end{widetext}

The above convention means that transferring a pair from terminal $1$ to the dot does not change $N$ while transfering this pair from the dot to terminal $2$ 
increases $N$ by $1$. This maps the IGM on a bipartite tight-binding chain, described by a model of the class of Rice-Mele models\cite{Rice-Mele}. The sites of this chain are indexed by $(N,\nu)$ and the phase $\varphi$ plays the role of the one-dimensional wavevector $k$ and both models yield the two-level Hamiltonian given by Equation \ref{twolevel}. The comparison with Rice-Mele model is more transparent in the limit $b<<1$ where the time dependence is harmonic and one has $H_\infty(t)= \vec{h}(t).\vec{\sigma}$ with

\begin{widetext}
\begin{equation}
\vec{h}(t)=\Big((\gamma_1+\gamma_2)\cos\frac{\varphi}{2}-(\gamma_1\delta\varphi_1(t)-\gamma_2\delta\varphi_2(t))\sin\frac{\varphi}{2},(\gamma_1-\gamma_2)\sin\frac{\varphi}{2}+(\gamma_1\delta\varphi_1(t)+\gamma_2\delta\varphi_2(t))\cos\frac{\varphi}{2},\varepsilon_0\Big)
\end{equation}
\end{widetext}
with $\delta\varphi_{1,2}(t)=b_{1,2}\cos(\omega t\pm\frac{\chi}{2})$. Notice that in absence of microwave excitation, the IGM maps onto a SSH model.

The existence of pumped charge current in such models, as proposed and realized in experiments\cite{pumping1,pumping2}, provides an interpretation of our results. An interesting point is that due to the presence of a continuum of quasiparticle states, the full model described by Equation \ref{twolevel}  goes well beyond such a simple chain model. This work shows that the pumping properties are robust against the incorporation of such continuum states. Moreover, pumping scenarios are usually considered in the quasi-adiabatic limit\cite{Thouless}, while here, we have studied the full frequency range and new resonant features.

\section{Conclusion}
In conclusion, this work demonstrates that two dephased microwave fields
microwaves provide nonadiabatic pumping of
Josephson currents without any superconducting phase difference. The
current is driven by the microwave phase $\chi$ and is tuned in
amplitude and sign by crossing ABS resonances. The chirality has its root in the structure of the wavefunction of the ABS
 as a function of two phases ($\varphi_1(t)$ and $\varphi_2(t)$, or equivalently $\varphi$ and $\Omega t$). The chiral properties
are robust against temperature {and asymmetry} in the
junction parameters ($\gamma_1\neq\gamma_2$) and in
the microwave amplitudes. Most results have been
shown with small microwave amplitudes, but larger values
of $I_{chir}$ comparable to $I_0$ can be easily be reached
in experiments. The striking sign change of the
current at resonance contrasts with the current amplitude minima found
in the nonchiral case\cite{Bergeret1,Bergeret2}. All the possible known regimes of current in a standard Josephson junction 
(harmonic or sawtooth phase dependence, ``$0$'' or ``$\pi$'' junction, as well as very anharmonic ones like in $\varphi_0$-junctions) 
are encountered for the 
chiral current as a function of $\chi$. Using an
electrostatic gate or tuning the microwave frequency offers a
fine control of the chiral current, in amplitude and sign. The latter
resulting from symmetry properties, Coulomb interactions are not
expected to change qualitatively the physics.

 In a Josephson transistor\cite{JTransistor}, the
current amplitude oscillates with the gate without any sign change as
the dot levels pass across the gap. Due to the additional gate-controlled sign change, the proposed set-up deserves the
  name of {\it chiral Josephson transistor}. A
  generalization to a multilevel dot is possible but involving also
  microwave transitions between different channel ABS.
  Also, the chiral current variation quadratic with the microwave amplitude recalls 
  the photogalvanic effect studied in Ref. \onlinecite{Grushin}.

 D. F.  gratefully acknowledges fruitful discussions with
 C. Balseiro and G. Usaj.
\bigskip
 \appendix
 \section{Rotating Wave Approximation}
 The general form of the  $\hat{H}_n(t)$ matrix in Equation (\ref{Hn}) is the following:
 \begin{widetext}
\begin{equation}
\begin{aligned}
\hat{H}_n(t)&=\hat{\tau}_z\left[E_A+\dot{\beta}\cos^2\frac{\theta}{2}-\frac{\dot{\beta}+\dot{\alpha}}{2}+\eta \sin\theta\pdv{|\Gamma|}{\varphi}-n\frac{\Omega}{2}\right]\\
&+\hat{\tau}_x\left[\eta \cos\theta\cos\alpha\pdv{|\Gamma|}{\varphi}\cos(n\Omega t)-\left(\frac{\dot{\theta}}{2\cos \alpha}+\eta \cos\theta\sin\alpha\pdv{|\Gamma|}{\varphi}\right)\sin(n\Omega t)\right]\\
&+\hat{\tau}_y\left[-\left(\frac{\dot{\theta}}{2\cos \alpha}+\eta \cos\theta\sin\alpha\pdv{|\Gamma|}{\varphi}\right)\cos(n\Omega t)-\eta \cos\theta\cos\alpha\pdv{|\Gamma|}{\varphi}\sin(n\Omega t)\right]\\
&+\frac{\dot{\beta}+\dot{\alpha}}{2}
.
\end{aligned}
\label{hamiltonian_rwa}
\end{equation}
\end{widetext}

We consider the cases $\varphi=0$ $(\zeta=1)$ and $\pi$ $(\zeta=-1)$:
\begin{align}
|\Gamma(t)|=\sqrt{2\gamma_0^2\left(1+\zeta\cos(2b\sin\Omega t \sin \frac{\chi}{2})\right)}
\label{Gamma}
\\ \pdv{|\Gamma(t)|}{\varphi}=-\frac{\zeta\gamma_0^2}{|\Gamma(t)|}\sin(2b\sin\Omega
\sin\frac{\chi}{2})\\ E_A(t)=\sqrt{\epsilon_0^2+|\Gamma(t)|^2}\\ \dot{\beta}=\Omega
b \sin\Omega t \cos
\frac{\chi}{2}\\ \cos[\theta(t)]=\frac{\epsilon_0}{E_A(t)}\\ \sin[\theta(t)]=\frac{|\Gamma(t)|}{E_A(t)}\\ \tan[\alpha(t)]=-\frac{\dot{\beta}\sin\theta}{\dot{\theta}}\label{tan_alpha}
.
\end{align}

Let us first consider $\varphi=0$, with $b$ small. Taylor expanding the
expression of $\tan\alpha$ in Eq. \eqref{tan_alpha} leads to
\begin{equation} 
(\tan[\alpha(t)])^{-1}=\frac{\epsilon_0}{E_0}b
\frac{\sin^2\frac{\chi}{2}}{\cos \frac{\chi}{2}}\cos\Omega t
,
\end{equation} 
meaning that $(\tan[\alpha(t)])^{-1}$ is not small if $\chi$ is
close to $\pi$. Imposing time-continuity of $\alpha$ implies $\sin
\alpha>0$. Therefore $\sin\alpha=\frac{1}{f(t)}$ where
$f(t)=\sqrt{1+\frac{\epsilon_0^2}{E_0^2} b^2 \cos^2\Omega t
  \frac{\sin^4\frac{\chi}{2}}{\cos^2 \frac{\chi}{2}}}$ and $\cos\alpha
=\frac{\epsilon_0}{E_0}b \frac{\sin^2\frac{\chi}{2}}{\cos
  \frac{\chi}{2}}\cos\Omega t/f(t)$. Using the first harmonic $n=1$,
Taylor expanding and integrating, Eq. \eqref{hamiltonian_parity}
yield Equation (\ref{Effective_H_0}) where $F_1, F_2$ are defined by
\begin{equation}
F_1 = \frac{\int_0^T\sin^2\Omega t f(t)}{\int_0^T\sin^2\Omega t}
\,,
\quad
F_2 = \frac{\int_0^T\sin^2\Omega t/f(t)}{\int_0^T\sin^2\Omega t}
.
\end{equation}

A similar calculation in the case $\varphi=\pi$ makes use of 

\begin{eqnarray}
\nonumber
G_1 = \frac{\int_0^T \left|\sin2\Omega t \right| g_1(t)}{\int_0^T \left|\sin2\Omega t \right|}
\,,
\quad
G_2 = \frac{\int_0^T \left|\sin\Omega t \right| \sin^2\Omega t  /g_2(t)}{\int_0^T \left|\sin\Omega t \right| \sin^2\Omega t}
,\\
\end{eqnarray}
with
\begin{equation}
g_1(t)=\sqrt{1+ b^2 \cos^2\frac{\chi}{2} \frac{\sin^4\Omega t}{\cos^2 \Omega t}}
\,,
\quad
 g_2(t)=\left|\cos\Omega t \right| g_1(t).
\end{equation}

 \section{Green's functions}
Let us define the bare Green's functions (GFs) and the tunnel self-energy for the calculation of the current.

\begin{equation}
\label{self-energy}
\Sigma_{jd}^{R,A}=
\begin{pmatrix}
t_{j}e^{i\varphi_j(t)/2}&0\\
0&-t_{j}e^{-i\varphi_j(t)/2}
\end{pmatrix}
\end{equation}

($j=1,2$). The bare GFs in the leads are given by

\begin{equation}
\hat{g}^{r,a}_{jj} = \frac{\pi \nu(0)} {\sqrt{\Delta^2 - (\omega \pm
    i\eta_s)^2}} 
 \begin{pmatrix} - (\omega \pm i\eta_s)& \Delta\\
 \Delta& - (\omega \pm i\eta_s)
 \end{pmatrix}
\end{equation}

and $\hat{g}^{+-}_{jj} (\omega) = n_F(\omega) \big (\hat{g}^{A}_{jj} ( \omega ) -
\hat{g}^{R}_{jj}(\omega)\big)$. 

The bare GFs in the dot are
  given by $\hat{g}^{R,A}_{dd}=(\omega-\varepsilon_0\hat{\tau}_z\pm
i\eta_d)^{-1}$ and $\hat{g}^{+-}_{dd}(\omega)=
n_F(\omega)\big(\hat{g}^{a}_{dd}(\omega)-\hat{g}^{r}_{dd}(\omega)\big)$. The
broadening parameters $\eta_{s},\eta_{d}$ mimic residual inelastic
processes.

Expanding the Dyson equation (\ref{Dyson}) yields
\begin{widetext}
\begin{equation}
\begin{aligned}
G_{d1}^{+-} =
&(1+G_{dd}^r\Sigma_{d1}^r g_{11}^r \Sigma_{1d}^r
+G_{dd}^r\Sigma_{d2}^r g_{22}^r\Sigma_{2d}^r)
g_{dd}^{+-}
(\Sigma_{d1}^a g_{11}^a + \Sigma_{d1}^a g_{11}^a\Sigma_{1d}^a G_{dd}^a \Sigma_{d1}^a g_{11}^a+\\ \Sigma_{d2}^a g_{22}^a\Sigma_{2d}^a G_{dd}^a \Sigma_{d1}^a g_{11}^a)
\\&+
G_{dd}^r\Sigma_{d1}^r g_{11}^{+-}(1+\Sigma_{1d}^a G_{dd}^a\Sigma_{d1}^a g_{11}^a)
+
G_{dd}^r\Sigma_{d2}^r g_{22}^{+-} \Sigma_{2d}^a G_{dd}^a\Sigma_{d1}^a g_{11}^a
\end{aligned}
\end{equation}
\begin{equation}
\begin{aligned}
G_{1d}^{+-} =
&(g_{11}^r \Sigma_{1d}^r + g_{11}^r\Sigma_{1d}^r G_{dd}^r \Sigma_{d1}^r g_{11}^r \Sigma_{1d}^r+g_{11}^r\Sigma_{1d}^r G_{dd}^r \Sigma_{d2}^r g_{22}^r \Sigma_{2d}^r)g_{dd}^{+-}(1+ \Sigma_{d1}^a g_{11}^a \Sigma_{1d}^a G_{dd}^a+\\ \Sigma_{d2}^a g_{22}^a \Sigma_{2d}^a G_{dd}^a)
\\&+
(1+g_{11}^r \Sigma_{1d}^r G_{dd}^r\Sigma_{d1}^r)g_{11}^{+-}\Sigma_{1d}^a G_{dd}^a
+
g_{11}^r \Sigma_{1d}^r G_{dd}^r \Sigma_{d2}^r g_{22}^{+-} \Sigma_{2d}^a G_{dd}^a
,
\end{aligned}
\end{equation}
\end{widetext}
where $G_{dd}^{r,a}$ is the full retarded or advanced Green's function localized on the dot such as
\begin{equation}
G_{dd}^{r,a}=((g_{dd}^{r,a})^{-1}-(\Sigma_{d1}g_{11}^{r,a}\Sigma_{1d}+\Sigma_{d2}g_{22}^{r,a}\Sigma_{2d}))^{-1}
.
\end{equation} 
Here $\Sigma_{dj,n}$
(respectively $\Sigma_{jd,n}$) is the $n$th harmonic of the
self-energy defined in Eq. \ref{self-energy}. 
\begin{widetext}
\begin{equation}
\begin{aligned}
\nonumber
\Sigma_{d1,n}=\Sigma^*_{1d,-n}=
\begin{pmatrix}
t_1e^{-i\varphi/4}(-i)^n J_n(\frac{b_1}{2})e^{in\chi/2} & 0 \\
0 & -t_1e^{i\varphi/4}i^n J_n(\frac{b_1}{2})e^{in\chi/2}
\end{pmatrix}
\end{aligned}
\end{equation}
\begin{equation}
\begin{aligned}
\Sigma_{d2,n}=\Sigma^*_{2d,-n}=
\begin{pmatrix}
t_2e^{i\varphi/4}(-i)^n J_n(\frac{b_2}{2})e^{-in\chi/2} & 0 \\
0 & -t_2e^{-i\varphi/4}i^n J_n(\frac{b_2}{2})e^{-in\chi/2}
\end{pmatrix}
.
\end{aligned}
\end{equation}
\end{widetext}

 \section{Crosstalk effects}
 
Let us consider the infinite gap model Hamiltonian with crosstalk
between the microwave amplitudes (and phases) applied on
superconductors $1,2$, and towards the electrostatic gate:
\begin{equation}
H_\infty(t)=
\begin{pmatrix}
\epsilon & \Gamma(t)\\
\Gamma^*(t) & -\epsilon
\end{pmatrix}
\label{Hamilonian_crosstalk}
,
\end{equation}
with $\Gamma(t)=\sum_{j=1,2}\gamma_je^{-i\varphi_j(t)}$, and:
\begin{align}
\nonumber
\varphi_1&=\frac{\varphi}{2}+b_1\cos(\Omega t+\frac{\chi}{2})+b_1^\prime\cos(\Omega t-\frac{\chi}{2})\\
\nonumber
\varphi_2&=-\frac{\varphi}{2}+b_2\cos(\Omega t-\frac{\chi}{2})+b_2^\prime\cos(\Omega t+\frac{\chi}{2})\\
\epsilon&=\epsilon_0+\epsilon_1\cos(\Omega t+\frac{\chi}{2})+\epsilon_2\cos(\Omega t-\frac{\chi}{2})
,
\end{align}
with $b_j,b_j^\prime,\epsilon_j>0$. The crosstalk from superconductor
$1$ to $2$ ($2$ to $1$) is described by the terms $b'_{1,2}$ and the
crosstalk with the gate voltage applied to the dot is described by
$\varepsilon_{1,2}$. Using only the first harmonics in the Fourier
decomposition of the Hamiltonian defined in
Eq. \eqref{Hamilonian_crosstalk} leads to
\begin{equation}
H_\infty(t)=\tilde{H}_{\infty 0}+\tilde{H}_{\infty 1}e^{i\Omega t}+\tilde{H}_{\infty -1}e^{-i\Omega t}
,
\end{equation}

Using Brillouin-Wigner perturbation theory, the effective
Hamiltonian becomes
\begin{equation}
 H_{\infty eff}= \tilde{H}_{\infty 0}+\frac{1}{\Omega}[\tilde{H}_{\infty 1},\tilde{H}_{\infty -1}]
,
\end{equation}
and after a straightforward calculation, its eigenvalues are found to be:
\begin{align}
\nonumber
E^2_\varphi= &\,\, \tilde{\epsilon}^2+\gamma_1^2+\gamma_2^2+2\gamma_1\gamma_2\cos\varphi\\\nonumber
&-\cos\chi\left[\gamma_1^2b_1b_1^\prime+\gamma_2^2b_2b_2^\prime+\gamma_1\gamma_2\cos\varphi\left(b_1 b_1^\prime+b_2 b_2^\prime\right)\right]\\\nonumber
&-\frac{2\sin\chi}{\Omega}\left[\gamma_1^2 B_1^2+ \gamma_2^2 B_2^2+\gamma_1\gamma_2\cos\varphi\left(B_1+B_2\right)\right]\\
&+\frac{\sin^2\chi}{\Omega^2}\left[\gamma_1^2B_1^2+\gamma_2^2B_2^2\right]
+2\frac{\gamma_1\gamma_2}{\Omega^2}\cos\varphi B_1B_2\sin^2\chi
\label{energy_crosstalk}
.
\end{align}
where
$\tilde{\epsilon}=\epsilon_0+\gamma_1\gamma_2\sin\varphi\sin\chi[b_1b_2-b_1^\prime
  b_2^\prime]$, $B_1=b_1\epsilon_2-b_1^\prime\epsilon_1$, and
$B_2=b_2^\prime\epsilon_2-b_2\epsilon_1$.
Thisleads to the
DC-current at $\varphi=0,\,\pi$:
\begin{equation}
I_{\varphi=0,\pi}=2e\frac{\gamma_1\gamma_2\epsilon_0\sin\chi\left(b_1b_2-b_1^\prime b_2^\prime\right)}{\Omega E_0^{eff}}
,
\end{equation}
where $E_0^{eff}=E_{\varphi=0,\pi}$ as given by (\ref{energy_crosstalk}).


\begin{thebibliography}{99}
\bibitem{footnote} This argument easily generalizes to the solution of the full model (Hamiltonian (\ref{Hamilt})), by oberving that  
the current is given by the derivative of the full free energy, including quasiparticles, and that in the adiabatic approximation, 
the latter is a function of the phase $\varphi_1(t)-\varphi_2(t)$ only.
\end{thebibliography}
\end{document}